%% file: bragg.tex
\documentstyle[preprint, amsmath, prl, aps]{revtex}

\begin{document}

\setcounter{figure}{0}

\draft

\title{How to measure the Bogoliubov quasiparticle amplitudes\\
in a trapped condensate}

\author{A. Brunello$^1$, F. Dalfovo$^2$, L. Pitaevskii$^{1,3}$, and 
S. Stringari$^1$}

\address{$^1$ Dipartimento di Fisica, Universit\`{a} di Trento, 
I-38050 Povo, Italy \\
and Istituto Nazionale per la Fisica della Materia, Unit\`{a} di Trento}
\address{$^2$ Dipartimento di Matematica e Fisica, Universit\`{a} 
Cattolica, Via Musei 41, Brescia, Italy \\
and Istituto Nazionale per la Fisica della Materia,
Gruppo collegato di Brescia}
\address{$^3$ Kapitza Institute for Physical Problems 
117334 Moscow, Russia}

\date{July 7, 2000}

\maketitle

\begin{abstract}
We propose an experiment, based on two consecutive Bragg pulses, to 
measure the momentum distribution of quasiparticle excitations in a 
trapped Bose gas at low temperature. With the first pulse one generates 
a bunch of excitations carrying momentum $q$, whose Doppler line is
measured by the second pulse. We show that this experiment can provide 
direct access to the amplitudes $u_{q}$ and $v_{q}$ characterizing the
Bogoliubov transformations from particles to quasiparticles. We simulate 
the behavior of the nonuniform gas by numerically solving the time dependent
Gross-Pitaevskii equation. 
\end{abstract}


More than 50 years ago Bogoliubov \cite{Bogoliubov47} developed the 
microscopic theory of interacting Bose gases. A crucial step of the theory 
is given by the so called Bogoliubov transformations  
\begin{eqnarray}
b_{{\bf q}} &=&u_{q}a_{{\bf q}}+v_{q}a_{-{\bf q}}^{\dagger }
\label{bogo_transf_1} \\
b_{{\bf q}}^{\dagger } &=&u_{q}a_{{\bf q}}^{\dagger }+v_{q}a_{-{\bf q}}
\label{bogo_transf_2}
\end{eqnarray}
which transform particle creation, $a$, and annihilation, $a^{\dagger }$,
operators into the corresponding quasiparticle operators $b$ and 
$b^{\dagger}$. The real coefficients $u_{q}$ and $v_{q}$ are 
known as quasiparticle amplitudes. The Bogoliubov transformations are the
combined effect of gauge symmetry breaking and of the interactions which are
responsible for the mixing between the particle creation and annihilation 
operators. In virtue of transformations 
(\ref{bogo_transf_1})-(\ref{bogo_transf_2}), the many-body Hamiltonian of 
the interacting Bose gas becomes diagonal in the $b_{{\bf q}}$'s, 
representing a system of free quasiparticles whose energy
is given by the famous Bogoliubov dispersion law: 
\begin{equation}
\epsilon (q)=\left[ q^{2}c^{2}+\left( \frac{q^{2}}{2m}\right)^{2} 
\right]^{1/2} \,\, .  \label{bogo_dispersion}
\end{equation}
In Eq.~(\ref{bogo_dispersion}), $c=[gn/m]^{1/2}$ is the sound velocity fixed
by the density of the gas, $n$, and by the parameter $g$ characterizing the
interaction term $g\sum\limits_{i<j}\delta ({\bf r}_{i}-{\bf r}_{j})$ of the
many-body Hamiltonian. The interaction parameter $g$ is determined by the 
s-wave scattering length $a$ through the relation $g=4 \pi \hbar^2 a/m$. 
The dispersion law (\ref{bogo_dispersion}) fixes the value of the
quasiparticle amplitudes $u_{q}$ and $v_{q}$, which can be written as 
\begin{eqnarray}
u_{q},v_{q} = \pm \frac{\epsilon(q) \pm q^{2}/{2m}}
{2 \sqrt{\epsilon (q) \ q^{2}/{2m}}} \, ;  
\label{bogo_v}
\end{eqnarray}
and satisfy the normalization condition $u_{q}^{2}-v_{q}^{2}=1$. At low
momentum transfer ($q^{2}/2m\ll mc^{2}$) the Bogoliubov excitations are
phonons characterized by the linear dispersion law $\epsilon =qc$ and the
amplitudes $u_{q}$ and $v_{q}$ exhibit the infrared divergence $u_{q}\sim
-v_{q}\sim (mc/2q)^{1/2}$. Vice-versa, at high momentum transfer the
dispersion law (\ref{bogo_dispersion}) approaches the free energy $q^{2}/2m$
and the Bogoliubov amplitudes take the ideal gas values $u_{q}=1$, $v_{q}=0$. 

Bogoliubov's theory has been developed also for nonuniform gases.
In this case, the dispersion law (\ref{bogo_dispersion}) can be 
defined locally through the density dependence of the sound velocity.
The theory has been successfully used to interpret the available experimental 
results on the propagation of phonons in trapped Bose-Einstein condensed
atomic gases, namely, the excitation of the lowest frequency modes 
\cite{JinEMWC96,MewesADKDK96}, corresponding to discretized phonon
oscillations of the system \cite{Stringari96,DalfovoGPS99}, the 
generation of wave packets propagating in the medium with the speed of 
sound \cite{Andrews97} and the excitation of phonons through inelastic 
photon scattering \cite{StamperCGIGPK99}. However, these experiments 
reveal the propagation of phonons only in coordinate 
space, where the equations of motion take the classical hydrodynamic form, 
and not in momentum space, where Bogoliubov's transformations 
(\ref{bogo_transf_1})-(\ref{bogo_transf_2}) exhibit their peculiar
character. 

In this work we suggest a procedure to measure the Bogoliubov parameters 
$u_{q}$ and $v_{q}$ in a trapped Bose-Einstein condensed gas. Our strategy is
based on the following two steps:

A) First, one generates a bunch of quasiparticles in the sample by means of
the technique already used in \cite{StamperCGIGPK99}. This is
based on an inelastic collisional process (two photon Bragg scattering)
which can be implemented with two detuned lasers transferring momentum 
${\bf q}$ and energy $\hbar \omega $ to the sample. Here ${\bf q}=\hbar 
({\bf k}_{1}-{\bf k}_{2})$ and $\omega =(\omega _{1}-\omega _{2})$ are fixed 
by the difference of the wave vectors and the corresponding frequencies of the 
two lasers. In order to excite quasiparticles in the phonon regime one should
satisfy the condition $q<mc$.\ Let us call $N_{ph}$ the
number of quasiparticles with momentum ${\bf q}$ generated by this first
Bragg pulse and let us assume, for simplicity, that the system can be
treated as a uniform gas. According to the Bogoliubov transformations 
(\ref{bogo_transf_1})-(\ref{bogo_transf_2}), the momentum distribution of 
the gas will be modified as 
\begin{equation}
n({\bf p})=n_{0}({\bf p})+N_{ph}\left( u_{q}^{2}\delta ({\bf p}-{\bf q})
+v_{q}^{2}\delta ({\bf p}+{\bf q})\right) \; ,  \label{bogo_n_p}
\end{equation}
where $n_{0}({\bf p})$ is the momentum distribution at equilibrium.
Equation~(\ref{bogo_n_p}) reveals the occurrence of two new terms describing 
particles propagating with directions parallel and antiparallel to the momentum 
${\bf q}$ of the quasi particles (hereafter called phonons) and weights
proportional, respectively, to $u_{q}^{2}$ and $v_{q}^{2}$. The total
momentum, ${\bf P}=\int d{\bf p}\ {\bf p} n({\bf p})$ carried by the system,
is equal to ${\bf q}N_{ph}$, as a result of the normalization condition 
$u_{q}^{2}-v_{q}^{2}=1$.

B) In the second step of the experiment one measures the momentum
distribution (\ref{bogo_n_p}) by sending a second Bragg pulse immediately 
after the first Bragg pulse. The momentum ${\bf Q}$, and the energy 
$\hbar \Omega$ transferred by the second pulse should be much larger than the 
ones of the first pulse since, in order to be sensitive to the momentum 
distribution of the sample, the scattering should probe the individual motion 
of particles \cite{StengerICSPK99,ZambelliPSS00}. 
More precisely, one must satisfy the condition $\hbar \Omega \sim
Q^{2}/2m\gg mc^{2}$. The measured quantity is
the dynamic structure factor which, in the large $Q$ regime, takes the form 
\cite{HohenbergP66} 
\begin{equation}
S(Q,\Omega )= \ \frac{m}{Q}\int
dp_{x}dp_{y}\ n(p_{x},p_{y},p_{z}) \label{dyn_str_fact}
\end{equation}
where $p_{z}=m(\hbar \Omega -Q^{2}/2m)/Q$ 
and we have assumed ${\bf Q}$ to be directed along the ${\bf z}$ axis. By 
inserting (\ref{bogo_n_p}) into (\ref{dyn_str_fact}), one finds that the 
dynamic structure factor exhibits, in addition to the original peak located at 
$\Omega =Q^{2}/2m\hbar $, two side peaks at 
\begin{equation}
\Omega _{\pm }=\frac{Q^{2}}{2m\hbar }\pm \frac{{\bf q}\cdot {\bf Q}}{m\hbar }
\; .  \label{bogo_energy_peaks_2}
\end{equation}
By denoting with $S_{+}$ and $S_{-}$ their contributions to the integrated
strength $\int d\Omega\ S(Q,\Omega )=N$, one finds $S_{+}=N_{ph}u_{q}^{2}$ and 
$S_{-}=N_{ph}v_{q}^{2}$ or, equivalently, $N_{ph}=S_{+}-S_{-}$ and 
$v_{q}^{2}=S_{-}/(S_{+}-S_{-})$. If the quantity $S_{+}+S_{-}=N_{ph}
(u_{q}^{2}+v_{q}^{2})$ is much smaller than $N$, the normalization of the
central peak remains close to the unperturbed value $N$. 
From the above discussion one concludes that the measurement of the dynamic
structure factor at high momentum transfer ${\bf Q}$ 
and, in particular, of the two strengths $S_{\pm }$ would
provide direct access to the number of phonons generated with the first
Bragg pulse, as well as to the value of the corresponding quasiparticle
amplitudes.

Expression (\ref{dyn_str_fact}) for the dynamic structure factor ignores the 
effects of the final state interactions which are responsible for both the 
line shift of the curve $S(Q,\Omega )$ and for its broadening. These effects 
can be safely calculated within Bogoliubov's theory  and, in the large $Q$ 
domain, are both fixed by the chemical potential of the gas 
\cite{StengerICSPK99,ZambelliPSS00}.
The broadening due to mean field effects should not be confused with the 
Doppler broadening included in Eq.~(\ref{dyn_str_fact}). The latter is due 
to the fact that, even in the equilibrium configuration,  the momentum 
distribution of the
condensate has a width $\sim \hbar /R_{z}$ originating from its zero point
motion in the ${\bf Q}$ direction. In the following, we will consider
condensates highly elongated along the axial $z$-axis so that the Doppler
broadening, due to the finite size of the system,
can be ignored. For a safe identification of the two
phonon peaks (\ref{bogo_energy_peaks_2}) and of the corresponding strengths 
$S_{\pm }$ it is crucial that the
separation $\Delta \Omega=\pm \left( {\bf q}\cdot {\bf Q}\right)/(m \hbar)$ 
between the phonon and central peaks be larger than the mean-field effect.
This imposes the condition 
\begin{equation}
q Q /m > \mu \; ,  \label{condition}
\end{equation}
where we have chosen the two vectors ${\bf q}$ and ${\bf Q}$ parallel in
order to maximize the separation $\Delta \Omega$. Equation~(\ref{condition}) 
shows that the momentum $q$ of the phonons generated by the first Bragg pulse 
should not be too small.

In the second part of the work we explore in detail the microscopic mechanisms 
of generation of phonons produced by the first Bragg pulse, taking into 
account the fact that our system is nonuniform and that the time duration 
of the pulse is finite. We consider a gas of interacting atoms initially 
confined by a harmonic potential of the form 
$V_{{\rm ho}}(x,y,z)=m\left( \omega _{\perp
}^{2}(x^{2}+y^{2})+\omega _{z}^{2}z^{2}\right)/2$.
The generation of phonons is analyzed through the numerical solution of the
time dependent Gross-Pitaevskii equation for the order parameter 
$\Psi({\bf r},t)$  \cite{DalfovoGPS99} in the presence of the additional 
external potential
\begin{equation}
V_{{\rm Bragg}}(z,t)=Vf(t)\left[\cos (q z/ \hbar -\omega t) \right] 
\label{Vbragg}
\end{equation}
which reproduces the effects of the inelastic scattering associated with the
two photon Bragg pulse directed along the axial $z$ direction 
(see for example Ref. \cite{BlakieB99}). In Eq.~(\ref{Vbragg}) 
the parameter $V$ is the strength of the Bragg pulse while the envelope
function $f(t)$ was chosen of the form
$f(t)=\frac{1}{2}\left[ 1+\tanh \left( t/t_{up} \right) \right]$,
and $f(t)=0$ for $t>t_B$. Here $t_B$ is the duration of the Bragg pulse,
while $t_{up}$ fixes its rise time.
By varying the values of $q$ and $\omega$ of the perturbation (\ref{Vbragg})
different regions of the nonuniform gas are excited with different 
intensity.  In fact, according to the
Bogoliubov dispersion (3), the condensate is
in resonance with the periodic perturbation $\cos (q z/ \hbar -\omega t)$ 
for values of the density satisfying the condition 
$c^2=gn/m=[\hbar^2 \omega^2-(q^2/2m)^2]/q^2$.

The ground state, corresponding to the stationary solution of the 
Gross-Pitaevskii equation at large negative times $t$, was obtained by 
means of the steepest descent method \cite{DalfovoS96}. For the time 
dependent solutions we have used a numerical code developed in 
Ref.~\cite{ModugnoD00}, suitable for axially symmetric condensates. The 
parameter $V$ has been chosen in order to generate a number of phonons 
corresponding to $5-10\%$ of the total number of atoms. In this way, one 
produces a visible bunch of excitations whose features can be still 
described using linear response theory. Higher values of $V$ were
also considered to explore nonlinear effects. The Bragg pulse
duration $t_B$ was always taken to be significantly less than
the oscillation time in the axial direction. This requirement is needed 
in order to relate the total momentum transferred by the photons with 
the actual momentum carried by the system at the end of the first Bragg 
pulse, thereby ignoring the effects of the external force produced by 
the harmonic potential during the pulse. This condition is well satisfied in
the experiment of Ref.~\cite{StamperCGIGPK99} where the total momentum $P_z$
of the condensate was measured after a Bragg pulse. 

An example of the density  $\left|\Psi({\bf r},t_B)\right|^2$ as a function 
of $z$ and for $r_\perp= [x^2+y^2]^{1/2}= 0$, is shown in Fig.~1. The 
condensate in this figure has $N=6\times10^7$ sodium atoms 
confined in a trap with $\omega_\perp =2 \pi \times 150$ Hz and 
$\omega_z = 0.12 \omega_\perp$. This corresponds to a Thomas-Fermi
parameter $Na/a_\perp = 10000$,
where $a_{\perp }=[\hbar /(m\omega _{\perp })]^{1/2}$. We have chosen a 
duration of the Bragg pulse $t_B=0.25 \times 2 \pi/\omega_\perp$ 
$(\sim1.7$ ms) and intensity $V=1.25 \hbar \omega_\perp$.  
The values of $q$ and $\omega$ are $q=1 \hbar / a_\perp$ 
and $\omega = 4.13 \omega_\perp$. With these parameters we are close to 
the phonon regime ($q^2/2m=0.02 \ mc^2$). 

In Fig.~2 we give the corresponding prediction for the dynamic structure
factor $S(Q,\Omega)$, measurable with the second Bragg pulse.
This quantity is evaluated in impulse approximation and is determined 
by the longitudinal momentum distribution, as in Eq.~(\ref{dyn_str_fact}),
\begin{eqnarray}
\int dp_{x}dp_{y}\ n(p_{x},p_{y},p_{z}) =  
& \int & dx' dy' dz' dz \ e^{-i p_z (z-z')/\hbar} \nonumber \\
&\times & \Psi^*(x',y',z,t_B) \Psi(x',y',z',t_B) \label{long_mom}
\end{eqnarray}
where $p_{z}=m(\hbar \Omega -Q^{2}/2m)/Q$. Final state interaction effects are 
ignored in this approximation, but they do not affect the conclusions 
of our analysis provided condition (\ref{condition}) is satisfied.

Figure 2 clearly shows the appearance of the two peaks in $S(Q,\Omega)$ 
at the frequencies predicted by Eq.~(\ref{bogo_energy_peaks_2}).  The 
difference $S_+ - S_-$ between their strengths gives the number of 
phonons $N_{ph}$; this turns out to be $\sim 6 \times 10^{6}$, 
i.e., about $10\%$ of the total number of atoms.
We have verified that the results are independent of the choice of the
rise time of the pulse, provided $t_{up} \le 0.05 \ 2 \pi / \omega_\perp$.
We have also checked that the system responds in a linear way, by verifying
that the value of $P_z$ increases quadratically with $V$,
and, for sufficiently long times, the number of phonons generated 
by the pulse increases linearly with $t_B$ as predicted by perturbation theory.
Moreover, we point out that condition (\ref{condition}), which ensures the 
visibility of the two peaks in $S(Q, \Omega)$, can be satisfied with
reasonable choices of the momentum $Q$ of the second Bragg pulse. Taking, 
for example, the value $Q=21 \mu m^{-1}$ \cite{StengerICSPK99} we get 
$Qq/m=36 \hbar \omega_\perp$,  to be compared with the value 
$\mu=25 \hbar \omega_\perp$ of the chemical potential. It is finally worth 
noticing that, since each phonon carries momentum $q$, their number 
can also be obtained by measuring the total momentum $P_z$ 
after the first Bragg pulse, as done in the experiment of 
Ref.~\cite{StamperCGIGPK99}: $N_{ph}=P_z/q$. This is useful when 
$q \ll mc$, since in this case 
$S_+ \sim S_-$ and the difference $S_+ - S_-$ may be difficult to extract. 

The strengths $S_{\pm }$ can be used to estimate the value of $v_{q}^{2}$.
Our results are given in Fig.~3 as a function of the first Bragg 
pulse duration. The three curves have been obtained with different
choices for the transferred energy and momentum, $\omega$ and $q$, but
they correspond to the same resonant density, i.e., the Bragg 
pulse excites the system in resonance at the same density ($\sim 0.67$
of the central value). The conditions for such resonant behavior have 
been taken from the local density approximation discussed 
in \cite{ZambelliPSS00}. Due to the different values of $q$, the three 
curves correspond also to different values of $v_q^2$, since this quantity 
depends on $q$ and on the density  through the ratio $mgn/q^2$, as predicted 
by Eqs.~(\ref{bogo_dispersion}-\ref{bogo_v}). Our results clearly
show this effect. In order to make the analysis more quantitative,
we also report, for each curve, the value predicted by 
Eq.~(\ref{bogo_v}) with $\epsilon(q) = \hbar \omega$. In the case of a 
periodic perturbation in a uniform gas the calculated curves of $v_q$ 
should coincide with prediction (\ref{bogo_v}).
In our calculations we find that the values of $v^2_q$ exhibit 
oscillations with frequency $2 \omega$ and a slight decrease as a
function of time. The behavior of $v^2_q$ at short times is the
consequence of the high frequency components contained in the Fourier 
transform of the Bragg potential (\ref{Vbragg}), whose effects cannot 
be simply  described employing a local density picture. The decrease of 
the signal at larger times is likely the consequence of the diffusion of 
phonons towards regions of lower density as well as of nonlinear effects. 
Despite these effects Fig.~3 clearly reveals the important features 
predicted by Bogoliubov theory for the quasiparticle amplitude $v$ and, in
particular, its dependence on the relevant parameters of the system. 

The results of Figs.~1-3 refer to conditions of linear or almost linear 
regime.  It is also interesting to explore the response of the condensate 
to a  highly nonlinear perturbation generating a number of excitations 
comparable to the total number of atoms. 
This can be achieved by increasing 
the strength $V$ of the Bragg pulse. In Fig.~4 we show the dynamic structure 
factor $S(Q,\Omega)$ calculated after the first Bragg pulse, in conditions 
of high nonlinearity ($V = 25 \hbar \omega_\perp$). Remarkably, the 
$p_z=0$ peak, corresponding to the initial  condensate, has almost 
disappeared. In this case, the appearance of additional peaks, associated 
with the second and third harmonics $p_z= \pm 2 q$ and $p_z= \pm 3 q$
in the longitudinal momentum distribution, is clearly visible.

In conclusion, we have suggested an experimental method to measure 
Bogoliubov's quasiparticle amplitudes in a trapped Bose gas at low 
temperature. In such an experiment the condensate is hit by a sequence 
of two Bragg pulses.  The first (low $q$ momentum transfer) pulse generates a 
bunch of phonons which are subsequently mapped in momentum space by the second 
(high $Q$) pulse. A 3D numerical simulation has allowed us to test our 
predictions and to show that our proposal is compatible with
the presently available experimental possibilities. This experiment
would provide the first direct measurement of the Bogoliubov quasiparticle
amplitudes,  which are of fundamental importance in the theory of Bose-Einstein 
condensation.

Useful conversations with M. Edwards, A.J. Leggett and W. Ketterle are 
acknowledged. 
A.B. would like to warmly thank M. Modugno for providing the code
needed for the simulations presented here. This work has been supported by the 
Ministero della Ricerca Scientifica e Tecnologica (MURST). F.D. thanks the
Dipartimento di Fisica dell'Universit\`{a} di Trento for the hospitality.


\begin{figure}
\begin{center}
\input{fig1.tex}
\caption{Density profile of the condensate as a function 
of $z$ evaluated at $r_\perp = 0$ after the first Bragg pulse.}
\label{Figure_1}
\end{center}
\end{figure}
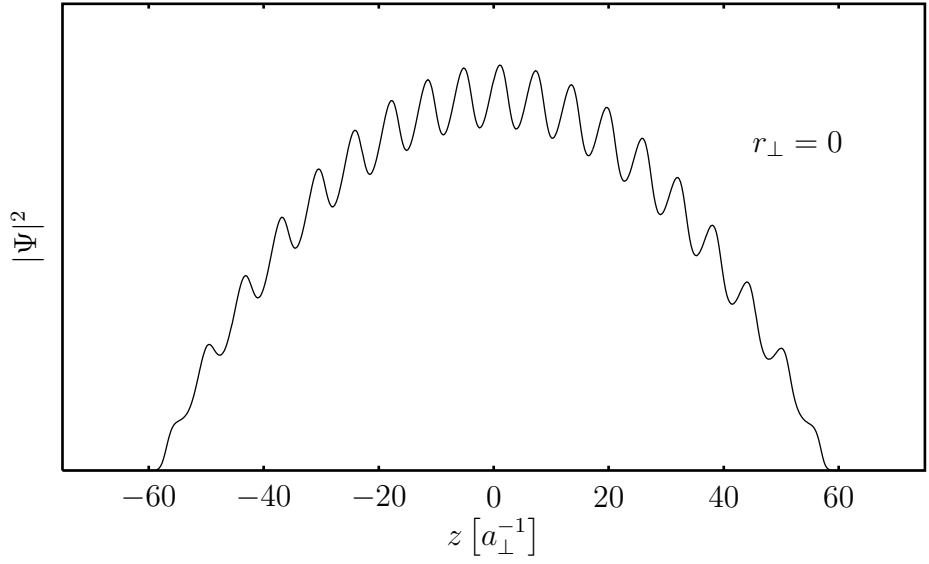
%
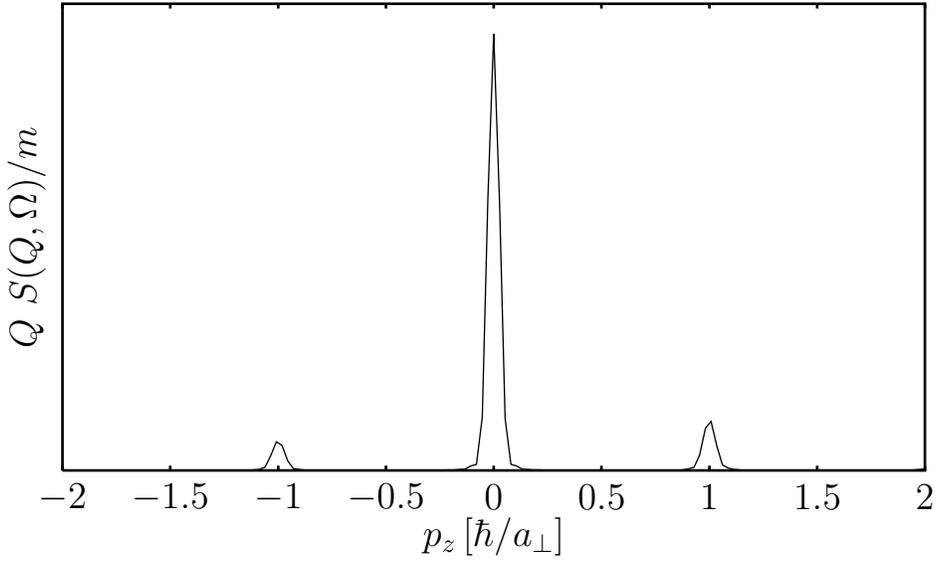
\begin{figure}
\begin{center}
\large
\input{fig2.tex}
\normalsize
\caption{Dynamic structure factor as a function of $p_z=m(\hbar \Omega
-Q^{2}/2m)/Q$ after the first Bragg pulse.}
\label{Figure_2}
\end{center}
\end{figure}
\begin{figure}
\begin{center}
\large
\input{fig3.tex}
\normalsize
\caption{$v_{q}^{2}$ as a function of time for different choices of 
$q$ (in units of $\hbar/a_\perp$) and $\omega$ (in units of $\hbar \omega_\perp$).
In order, from top to bottom: 
$q=1$, $\omega = 4.13$; $q=1.5$, $\omega = 6.25$ and $q=2$, $\omega = 8.44$. 
The values predicted by Eq.~(\ref{bogo_v}) 
are also reported.}
\label{Figure_3}
\end{center}
\end{figure}
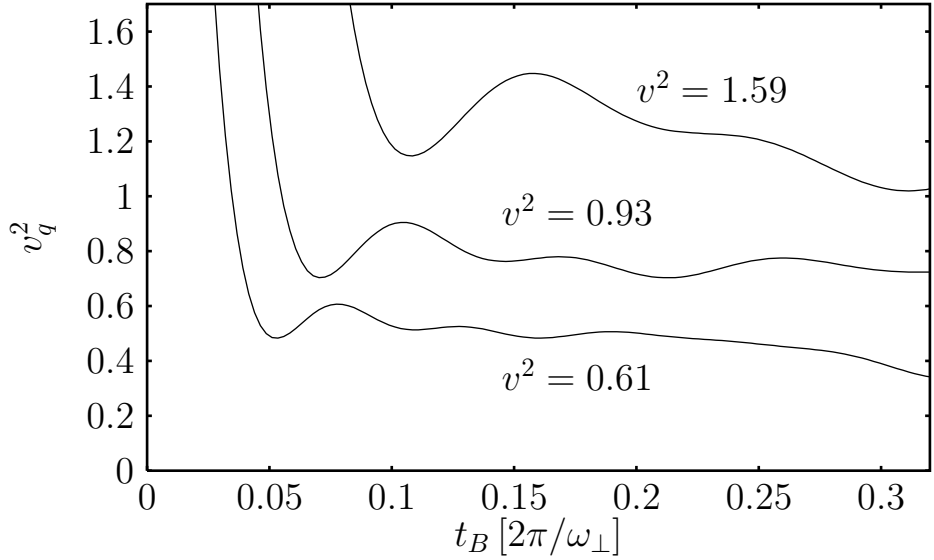
\begin{figure}
\begin{center}
\large
\input{fig4.tex}
\normalsize
\caption{Same as Figure 2. Here V is large 
in order to drive the system out of the linear regime.}
\label{Figure_4}
\end{center}
\end{figure}
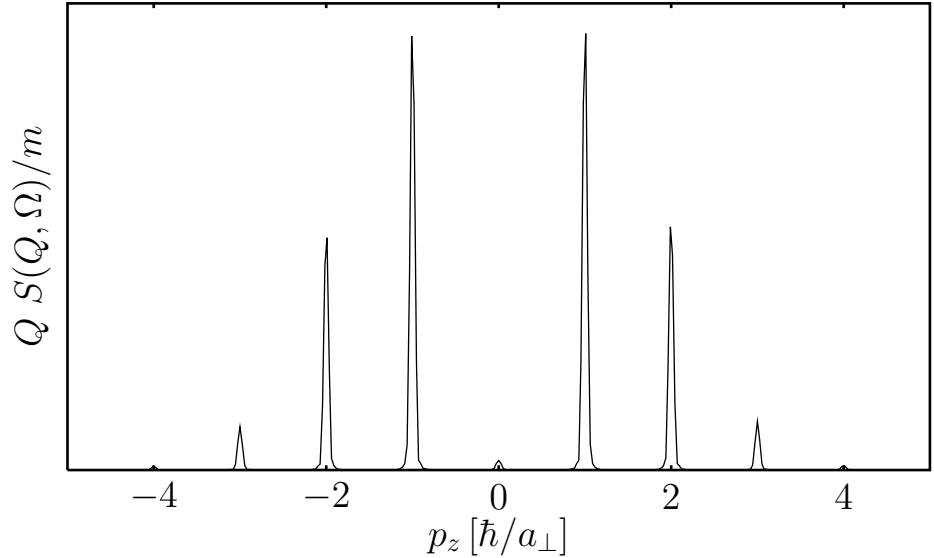

\end{document}

%% file: fig1.tex
\begingroup%
  \makeatletter%
  \newcommand{\GNUPLOTspecial}{%
    \@sanitize\catcode`\%=14\relax\special}%
  \setlength{\unitlength}{0.1bp}%
{\GNUPLOTspecial{!
/gnudict 256 dict def
gnudict begin
/Color false def
/Solid false def
/gnulinewidth 5.000 def
/userlinewidth gnulinewidth def
/vshift -33 def
/dl {10 mul} def
/hpt_ 31.5 def
/vpt_ 31.5 def
/hpt hpt_ def
/vpt vpt_ def
/M {moveto} bind def
/L {lineto} bind def
/R {rmoveto} bind def
/V {rlineto} bind def
/vpt2 vpt 2 mul def
/hpt2 hpt 2 mul def
/Lshow { currentpoint stroke M
  0 vshift R show } def
/Rshow { currentpoint stroke M
  dup stringwidth pop neg vshift R show } def
/Cshow { currentpoint stroke M
  dup stringwidth pop -2 div vshift R show } def
/UP { dup vpt_ mul /vpt exch def hpt_ mul /hpt exch def
  /hpt2 hpt 2 mul def /vpt2 vpt 2 mul def } def
/DL { Color {setrgbcolor Solid {pop []} if 0 setdash }
 {pop pop pop Solid {pop []} if 0 setdash} ifelse } def
/BL { stroke userlinewidth 2 mul setlinewidth } def
/AL { stroke userlinewidth 2 div setlinewidth } def
/UL { dup gnulinewidth mul /userlinewidth exch def
      10 mul /udl exch def } def
/PL { stroke userlinewidth setlinewidth } def
/LTb { BL [] 0 0 0 DL } def
/LTa { AL [1 udl mul 2 udl mul] 0 setdash 0 0 0 setrgbcolor } def
/LT0 { PL [] 1 0 0 DL } def
/LT1 { PL [4 dl 2 dl] 0 1 0 DL } def
/LT2 { PL [2 dl 3 dl] 0 0 1 DL } def
/LT3 { PL [1 dl 1.5 dl] 1 0 1 DL } def
/LT4 { PL [5 dl 2 dl 1 dl 2 dl] 0 1 1 DL } def
/LT5 { PL [4 dl 3 dl 1 dl 3 dl] 1 1 0 DL } def
/LT6 { PL [2 dl 2 dl 2 dl 4 dl] 0 0 0 DL } def
/LT7 { PL [2 dl 2 dl 2 dl 2 dl 2 dl 4 dl] 1 0.3 0 DL } def
/LT8 { PL [2 dl 2 dl 2 dl 2 dl 2 dl 2 dl 2 dl 4 dl] 0.5 0.5 0.5 DL } def
/Pnt { stroke [] 0 setdash
   gsave 1 setlinecap M 0 0 V stroke grestore } def
/Dia { stroke [] 0 setdash 2 copy vpt add M
  hpt neg vpt neg V hpt vpt neg V
  hpt vpt V hpt neg vpt V closepath stroke
  Pnt } def
/Pls { stroke [] 0 setdash vpt sub M 0 vpt2 V
  currentpoint stroke M
  hpt neg vpt neg R hpt2 0 V stroke
  } def
/Box { stroke [] 0 setdash 2 copy exch hpt sub exch vpt add M
  0 vpt2 neg V hpt2 0 V 0 vpt2 V
  hpt2 neg 0 V closepath stroke
  Pnt } def
/Crs { stroke [] 0 setdash exch hpt sub exch vpt add M
  hpt2 vpt2 neg V currentpoint stroke M
  hpt2 neg 0 R hpt2 vpt2 V stroke } def
/TriU { stroke [] 0 setdash 2 copy vpt 1.12 mul add M
  hpt neg vpt -1.62 mul V
  hpt 2 mul 0 V
  hpt neg vpt 1.62 mul V closepath stroke
  Pnt  } def
/Star { 2 copy Pls Crs } def
/BoxF { stroke [] 0 setdash exch hpt sub exch vpt add M
  0 vpt2 neg V  hpt2 0 V  0 vpt2 V
  hpt2 neg 0 V  closepath fill } def
/TriUF { stroke [] 0 setdash vpt 1.12 mul add M
  hpt neg vpt -1.62 mul V
  hpt 2 mul 0 V
  hpt neg vpt 1.62 mul V closepath fill } def
/TriD { stroke [] 0 setdash 2 copy vpt 1.12 mul sub M
  hpt neg vpt 1.62 mul V
  hpt 2 mul 0 V
  hpt neg vpt -1.62 mul V closepath stroke
  Pnt  } def
/TriDF { stroke [] 0 setdash vpt 1.12 mul sub M
  hpt neg vpt 1.62 mul V
  hpt 2 mul 0 V
  hpt neg vpt -1.62 mul V closepath fill} def
/DiaF { stroke [] 0 setdash vpt add M
  hpt neg vpt neg V hpt vpt neg V
  hpt vpt V hpt neg vpt V closepath fill } def
/Pent { stroke [] 0 setdash 2 copy gsave
  translate 0 hpt M 4 {72 rotate 0 hpt L} repeat
  closepath stroke grestore Pnt } def
/PentF { stroke [] 0 setdash gsave
  translate 0 hpt M 4 {72 rotate 0 hpt L} repeat
  closepath fill grestore } def
/Circle { stroke [] 0 setdash 2 copy
  hpt 0 360 arc stroke Pnt } def
/CircleF { stroke [] 0 setdash hpt 0 360 arc fill } def
/C0 { BL [] 0 setdash 2 copy moveto vpt 90 450  arc } bind def
/C1 { BL [] 0 setdash 2 copy        moveto
       2 copy  vpt 0 90 arc closepath fill
               vpt 0 360 arc closepath } bind def
/C2 { BL [] 0 setdash 2 copy moveto
       2 copy  vpt 90 180 arc closepath fill
               vpt 0 360 arc closepath } bind def
/C3 { BL [] 0 setdash 2 copy moveto
       2 copy  vpt 0 180 arc closepath fill
               vpt 0 360 arc closepath } bind def
/C4 { BL [] 0 setdash 2 copy moveto
       2 copy  vpt 180 270 arc closepath fill
               vpt 0 360 arc closepath } bind def
/C5 { BL [] 0 setdash 2 copy moveto
       2 copy  vpt 0 90 arc
       2 copy moveto
       2 copy  vpt 180 270 arc closepath fill
               vpt 0 360 arc } bind def
/C6 { BL [] 0 setdash 2 copy moveto
      2 copy  vpt 90 270 arc closepath fill
              vpt 0 360 arc closepath } bind def
/C7 { BL [] 0 setdash 2 copy moveto
      2 copy  vpt 0 270 arc closepath fill
              vpt 0 360 arc closepath } bind def
/C8 { BL [] 0 setdash 2 copy moveto
      2 copy vpt 270 360 arc closepath fill
              vpt 0 360 arc closepath } bind def
/C9 { BL [] 0 setdash 2 copy moveto
      2 copy  vpt 270 450 arc closepath fill
              vpt 0 360 arc closepath } bind def
/C10 { BL [] 0 setdash 2 copy 2 copy moveto vpt 270 360 arc closepath fill
       2 copy moveto
       2 copy vpt 90 180 arc closepath fill
               vpt 0 360 arc closepath } bind def
/C11 { BL [] 0 setdash 2 copy moveto
       2 copy  vpt 0 180 arc closepath fill
       2 copy moveto
       2 copy  vpt 270 360 arc closepath fill
               vpt 0 360 arc closepath } bind def
/C12 { BL [] 0 setdash 2 copy moveto
       2 copy  vpt 180 360 arc closepath fill
               vpt 0 360 arc closepath } bind def
/C13 { BL [] 0 setdash  2 copy moveto
       2 copy  vpt 0 90 arc closepath fill
       2 copy moveto
       2 copy  vpt 180 360 arc closepath fill
               vpt 0 360 arc closepath } bind def
/C14 { BL [] 0 setdash 2 copy moveto
       2 copy  vpt 90 360 arc closepath fill
               vpt 0 360 arc } bind def
/C15 { BL [] 0 setdash 2 copy vpt 0 360 arc closepath fill
               vpt 0 360 arc closepath } bind def
/Rec   { newpath 4 2 roll moveto 1 index 0 rlineto 0 exch rlineto
       neg 0 rlineto closepath } bind def
/Square { dup Rec } bind def
/Bsquare { vpt sub exch vpt sub exch vpt2 Square } bind def
/S0 { BL [] 0 setdash 2 copy moveto 0 vpt rlineto BL Bsquare } bind def
/S1 { BL [] 0 setdash 2 copy vpt Square fill Bsquare } bind def
/S2 { BL [] 0 setdash 2 copy exch vpt sub exch vpt Square fill Bsquare } bind def
/S3 { BL [] 0 setdash 2 copy exch vpt sub exch vpt2 vpt Rec fill Bsquare } bind def
/S4 { BL [] 0 setdash 2 copy exch vpt sub exch vpt sub vpt Square fill Bsquare } bind def
/S5 { BL [] 0 setdash 2 copy 2 copy vpt Square fill
       exch vpt sub exch vpt sub vpt Square fill Bsquare } bind def
/S6 { BL [] 0 setdash 2 copy exch vpt sub exch vpt sub vpt vpt2 Rec fill Bsquare } bind def
/S7 { BL [] 0 setdash 2 copy exch vpt sub exch vpt sub vpt vpt2 Rec fill
       2 copy vpt Square fill
       Bsquare } bind def
/S8 { BL [] 0 setdash 2 copy vpt sub vpt Square fill Bsquare } bind def
/S9 { BL [] 0 setdash 2 copy vpt sub vpt vpt2 Rec fill Bsquare } bind def
/S10 { BL [] 0 setdash 2 copy vpt sub vpt Square fill 2 copy exch vpt sub exch vpt Square fill
       Bsquare } bind def
/S11 { BL [] 0 setdash 2 copy vpt sub vpt Square fill 2 copy exch vpt sub exch vpt2 vpt Rec fill
       Bsquare } bind def
/S12 { BL [] 0 setdash 2 copy exch vpt sub exch vpt sub vpt2 vpt Rec fill Bsquare } bind def
/S13 { BL [] 0 setdash 2 copy exch vpt sub exch vpt sub vpt2 vpt Rec fill
       2 copy vpt Square fill Bsquare } bind def
/S14 { BL [] 0 setdash 2 copy exch vpt sub exch vpt sub vpt2 vpt Rec fill
       2 copy exch vpt sub exch vpt Square fill Bsquare } bind def
/S15 { BL [] 0 setdash 2 copy Bsquare fill Bsquare } bind def
/D0 { gsave translate 45 rotate 0 0 S0 stroke grestore } bind def
/D1 { gsave translate 45 rotate 0 0 S1 stroke grestore } bind def
/D2 { gsave translate 45 rotate 0 0 S2 stroke grestore } bind def
/D3 { gsave translate 45 rotate 0 0 S3 stroke grestore } bind def
/D4 { gsave translate 45 rotate 0 0 S4 stroke grestore } bind def
/D5 { gsave translate 45 rotate 0 0 S5 stroke grestore } bind def
/D6 { gsave translate 45 rotate 0 0 S6 stroke grestore } bind def
/D7 { gsave translate 45 rotate 0 0 S7 stroke grestore } bind def
/D8 { gsave translate 45 rotate 0 0 S8 stroke grestore } bind def
/D9 { gsave translate 45 rotate 0 0 S9 stroke grestore } bind def
/D10 { gsave translate 45 rotate 0 0 S10 stroke grestore } bind def
/D11 { gsave translate 45 rotate 0 0 S11 stroke grestore } bind def
/D12 { gsave translate 45 rotate 0 0 S12 stroke grestore } bind def
/D13 { gsave translate 45 rotate 0 0 S13 stroke grestore } bind def
/D14 { gsave translate 45 rotate 0 0 S14 stroke grestore } bind def
/D15 { gsave translate 45 rotate 0 0 S15 stroke grestore } bind def
/DiaE { stroke [] 0 setdash vpt add M
  hpt neg vpt neg V hpt vpt neg V
  hpt vpt V hpt neg vpt V closepath stroke } def
/BoxE { stroke [] 0 setdash exch hpt sub exch vpt add M
  0 vpt2 neg V hpt2 0 V 0 vpt2 V
  hpt2 neg 0 V closepath stroke } def
/TriUE { stroke [] 0 setdash vpt 1.12 mul add M
  hpt neg vpt -1.62 mul V
  hpt 2 mul 0 V
  hpt neg vpt 1.62 mul V closepath stroke } def
/TriDE { stroke [] 0 setdash vpt 1.12 mul sub M
  hpt neg vpt 1.62 mul V
  hpt 2 mul 0 V
  hpt neg vpt -1.62 mul V closepath stroke } def
/PentE { stroke [] 0 setdash gsave
  translate 0 hpt M 4 {72 rotate 0 hpt L} repeat
  closepath stroke grestore } def
/CircE { stroke [] 0 setdash 
  hpt 0 360 arc stroke } def
/Opaque { gsave closepath 1 setgray fill grestore 0 setgray closepath } def
/DiaW { stroke [] 0 setdash vpt add M
  hpt neg vpt neg V hpt vpt neg V
  hpt vpt V hpt neg vpt V Opaque stroke } def
/BoxW { stroke [] 0 setdash exch hpt sub exch vpt add M
  0 vpt2 neg V hpt2 0 V 0 vpt2 V
  hpt2 neg 0 V Opaque stroke } def
/TriUW { stroke [] 0 setdash vpt 1.12 mul add M
  hpt neg vpt -1.62 mul V
  hpt 2 mul 0 V
  hpt neg vpt 1.62 mul V Opaque stroke } def
/TriDW { stroke [] 0 setdash vpt 1.12 mul sub M
  hpt neg vpt 1.62 mul V
  hpt 2 mul 0 V
  hpt neg vpt -1.62 mul V Opaque stroke } def
/PentW { stroke [] 0 setdash gsave
  translate 0 hpt M 4 {72 rotate 0 hpt L} repeat
  Opaque stroke grestore } def
/CircW { stroke [] 0 setdash 
  hpt 0 360 arc Opaque stroke } def
/BoxFill { gsave Rec 1 setgray fill grestore } def
end
}}%
\begin{picture}(3600,2160)(0,-100)%
{\GNUPLOTspecial{"
gnudict begin
gsave
0 0 translate
0.100 0.100 scale
0 setgray
newpath
1.000 UL
LTb
525 300 M
0 18 V
0 1742 R
0 -18 V
958 300 M
0 18 V
0 1742 R
0 -18 V
1392 300 M
0 18 V
0 1742 R
0 -18 V
1825 300 M
0 18 V
0 1742 R
0 -18 V
2258 300 M
0 18 V
0 1742 R
0 -18 V
2692 300 M
0 18 V
0 1742 R
0 -18 V
3125 300 M
0 18 V
0 1742 R
0 -18 V
1.000 UL
LTb
200 300 M
3250 0 V
0 1760 V
-3250 0 V
200 300 L
1.000 UL
LT0
200 300 M
4 0 V
5 0 V
5 0 V
5 0 V
5 0 V
5 0 V
5 0 V
5 0 V
5 0 V
5 0 V
5 0 V
5 0 V
5 0 V
5 0 V
5 0 V
5 0 V
5 0 V
5 0 V
5 0 V
5 0 V
5 0 V
5 0 V
5 0 V
5 0 V
5 0 V
5 0 V
5 0 V
5 0 V
6 0 V
5 0 V
5 0 V
5 0 V
5 0 V
5 0 V
5 0 V
5 0 V
5 0 V
5 0 V
5 0 V
5 0 V
5 0 V
5 0 V
5 0 V
5 0 V
5 0 V
5 0 V
5 0 V
5 0 V
5 0 V
5 0 V
5 0 V
5 0 V
5 0 V
5 0 V
5 0 V
5 0 V
5 0 V
5 0 V
5 0 V
5 0 V
5 0 V
5 0 V
5 0 V
5 0 V
5 0 V
5 0 V
5 0 V
5 0 V
5 1 V
5 0 V
5 2 V
5 2 V
5 5 V
5 7 V
5 10 V
5 15 V
5 18 V
5 20 V
5 21 V
5 20 V
5 17 V
5 15 V
5 11 V
5 8 V
5 6 V
5 4 V
5 3 V
5 3 V
5 3 V
5 3 V
5 4 V
5 5 V
5 6 V
5 7 V
5 10 V
5 11 V
5 13 V
5 16 V
5 17 V
5 20 V
5 21 V
5 22 V
5 23 V
5 22 V
5 22 V
5 19 V
5 16 V
5 14 V
5 9 V
5 4 V
5 0 V
5 -3 V
5 -6 V
5 -8 V
5 -8 V
5 -7 V
5 -5 V
5 -3 V
5 0 V
5 3 V
5 6 V
5 8 V
5 12 V
5 15 V
5 17 V
5 20 V
5 21 V
6 23 V
5 25 V
5 24 V
5 25 V
5 23 V
5 22 V
5 19 V
5 16 V
5 12 V
5 7 V
5 2 V
5 -3 V
5 -9 V
5 -12 V
5 -14 V
5 -14 V
5 -13 V
5 -10 V
5 -7 V
5 -2 V
5 2 V
5 5 V
5 10 V
5 13 V
5 16 V
5 19 V
5 21 V
5 23 V
5 25 V
5 25 V
5 25 V
5 25 V
5 24 V
5 21 V
5 19 V
5 15 V
5 11 V
5 5 V
5 -1 V
5 -7 V
5 -12 V
5 -16 V
5 -19 V
5 -19 V
5 -17 V
5 -13 V
5 -9 V
5 -4 V
5 2 V
5 5 V
5 10 V
5 14 V
5 16 V
5 20 V
5 22 V
5 24 V
5 25 V
5 25 V
5 26 V
5 24 V
5 24 V
5 21 V
5 17 V
5 14 V
5 8 V
5 2 V
5 -3 V
5 -10 V
5 -16 V
5 -20 V
5 -23 V
5 -22 V
5 -19 V
5 -15 V
5 -10 V
5 -5 V
5 1 V
5 6 V
5 10 V
5 14 V
5 17 V
5 21 V
5 22 V
5 24 V
5 25 V
5 26 V
5 25 V
5 25 V
5 22 V
5 19 V
5 16 V
5 11 V
5 5 V
5 -1 V
5 -7 V
5 -14 V
5 -20 V
5 -24 V
6 -25 V
5 -24 V
5 -21 V
5 -16 V
5 -10 V
5 -4 V
5 1 V
5 7 V
5 11 V
5 15 V
5 18 V
5 21 V
5 23 V
5 24 V
5 25 V
5 26 V
5 24 V
5 24 V
5 20 V
5 18 V
5 12 V
5 8 V
5 1 V
5 -5 V
5 -12 V
5 -19 V
5 -24 V
5 -27 V
5 -28 V
5 -25 V
5 -21 V
5 -15 V
5 -9 V
5 -3 V
5 3 V
5 8 V
5 12 V
5 16 V
5 19 V
5 22 V
5 23 V
5 25 V
5 25 V
5 24 V
5 24 V
5 21 V
5 18 V
5 14 V
5 9 V
5 4 V
5 -4 V
5 -11 V
5 -17 V
5 -24 V
5 -28 V
5 -30 V
5 -29 V
5 -25 V
5 -20 V
5 -13 V
5 -6 V
5 -1 V
5 5 V
5 9 V
5 14 V
5 18 V
5 20 V
5 22 V
5 24 V
5 24 V
5 24 V
5 24 V
5 21 V
5 19 V
5 14 V
5 10 V
5 4 V
5 -3 V
5 -9 V
5 -17 V
5 -24 V
5 -29 V
5 -31 V
5 -32 V
5 -28 V
5 -23 V
5 -16 V
5 -10 V
5 -4 V
5 3 V
5 7 V
5 12 V
5 15 V
5 19 V
5 21 V
5 23 V
5 23 V
5 24 V
5 22 V
6 21 V
5 19 V
5 14 V
5 10 V
5 4 V
5 -2 V
5 -10 V
5 -16 V
5 -24 V
5 -30 V
5 -32 V
5 -34 V
5 -30 V
5 -26 V
5 -19 V
5 -12 V
5 -6 V
5 1 V
5 5 V
5 10 V
5 14 V
5 18 V
5 19 V
5 22 V
5 22 V
5 23 V
5 21 V
5 21 V
5 17 V
5 14 V
5 9 V
5 4 V
5 -3 V
5 -10 V
5 -17 V
5 -24 V
5 -31 V
5 -34 V
5 -34 V
5 -32 V
5 -27 V
5 -21 V
5 -14 V
5 -7 V
5 -1 V
5 4 V
5 9 V
5 12 V
5 16 V
5 19 V
5 20 V
5 21 V
5 21 V
5 21 V
5 19 V
5 16 V
5 13 V
5 8 V
5 3 V
5 -4 V
5 -11 V
5 -19 V
5 -25 V
5 -32 V
5 -35 V
5 -36 V
5 -33 V
5 -28 V
5 -21 V
5 -14 V
5 -8 V
5 -2 V
5 3 V
5 8 V
5 12 V
currentpoint stroke M
5 14 V
5 17 V
5 19 V
5 20 V
5 20 V
5 19 V
5 17 V
5 15 V
5 11 V
5 7 V
5 0 V
5 -5 V
5 -13 V
5 -21 V
5 -27 V
5 -33 V
5 -37 V
5 -36 V
5 -33 V
5 -28 V
5 -21 V
5 -14 V
5 -8 V
5 -2 V
6 3 V
5 7 V
5 11 V
5 13 V
5 16 V
5 18 V
5 18 V
5 18 V
5 17 V
5 16 V
5 12 V
5 9 V
5 4 V
5 -2 V
5 -8 V
5 -16 V
5 -23 V
5 -29 V
5 -35 V
5 -37 V
5 -37 V
5 -32 V
5 -26 V
5 -20 V
5 -14 V
5 -7 V
5 -2 V
5 3 V
5 6 V
5 10 V
5 13 V
5 15 V
5 15 V
5 17 V
5 16 V
5 15 V
5 13 V
5 10 V
5 6 V
5 1 V
5 -5 V
5 -12 V
5 -19 V
5 -26 V
5 -32 V
5 -36 V
5 -37 V
5 -35 V
5 -31 V
5 -24 V
5 -18 V
5 -12 V
5 -6 V
5 -2 V
5 3 V
5 6 V
5 9 V
5 11 V
5 13 V
5 14 V
5 14 V
5 13 V
5 13 V
5 10 V
5 7 V
5 2 V
5 -3 V
5 -9 V
5 -15 V
5 -23 V
5 -30 V
5 -34 V
5 -36 V
5 -35 V
5 -32 V
5 -27 V
5 -21 V
5 -15 V
5 -10 V
5 -6 V
5 -2 V
5 2 V
5 5 V
5 7 V
5 9 V
5 11 V
5 11 V
5 11 V
5 10 V
5 9 V
5 6 V
5 3 V
5 -2 V
5 -7 V
5 -14 V
5 -21 V
5 -26 V
5 -31 V
6 -34 V
5 -33 V
5 -30 V
5 -26 V
5 -21 V
5 -17 V
5 -12 V
5 -9 V
5 -6 V
5 -2 V
5 0 V
5 2 V
5 4 V
5 5 V
5 7 V
5 6 V
5 7 V
5 6 V
5 3 V
5 1 V
5 -3 V
5 -8 V
5 -13 V
5 -19 V
5 -24 V
5 -27 V
5 -29 V
5 -27 V
5 -25 V
5 -21 V
5 -17 V
5 -15 V
5 -12 V
5 -10 V
5 -8 V
5 -7 V
5 -5 V
5 -5 V
5 -3 V
5 -4 V
5 -3 V
5 -3 V
5 -4 V
5 -5 V
5 -8 V
5 -10 V
5 -14 V
5 -17 V
5 -21 V
5 -22 V
5 -22 V
5 -18 V
5 -13 V
5 -8 V
5 -5 V
5 -3 V
5 -2 V
5 0 V
5 -1 V
5 0 V
5 0 V
5 0 V
5 0 V
5 0 V
5 0 V
5 0 V
5 0 V
5 0 V
5 0 V
5 0 V
5 0 V
5 0 V
5 0 V
5 0 V
5 0 V
5 0 V
5 0 V
5 0 V
5 0 V
5 0 V
5 0 V
5 0 V
5 0 V
5 0 V
5 0 V
5 0 V
5 0 V
5 0 V
5 0 V
5 0 V
5 0 V
5 0 V
5 0 V
5 0 V
5 0 V
5 0 V
5 0 V
5 0 V
5 0 V
6 0 V
5 0 V
5 0 V
5 0 V
5 0 V
5 0 V
5 0 V
5 0 V
5 0 V
5 0 V
5 0 V
5 0 V
5 0 V
5 0 V
5 0 V
5 0 V
5 0 V
5 0 V
5 0 V
5 0 V
5 0 V
5 0 V
5 0 V
5 0 V
5 0 V
5 0 V
5 0 V
5 0 V
4 0 V
stroke
grestore
end
showpage
}}%
\put(2800,1532){\makebox(0,0)[l]{$r_\perp = 0$}}%
\put(1825,50){\makebox(0,0){$z \left[ a^{-1}_\perp \right]$}}%
\put(100,1180){%
\special{ps: gsave currentpoint currentpoint translate
270 rotate neg exch neg exch translate}%
\makebox(0,0)[b]{\shortstack{$|\Psi|^2$}}%
\special{ps: currentpoint grestore moveto}%
}%
\put(3125,200){\makebox(0,0){$60$}}%
\put(2692,200){\makebox(0,0){$40$}}%
\put(2258,200){\makebox(0,0){$20$}}%
\put(1825,200){\makebox(0,0){$0$}}%
\put(1392,200){\makebox(0,0){$-20$}}%
\put(958,200){\makebox(0,0){$-40$}}%
\put(525,200){\makebox(0,0){$-60$}}%
\end{picture}%
\endgroup
 

%% file: fig2.tex
\begingroup%
  \makeatletter%
  \newcommand{\GNUPLOTspecial}{%
    \@sanitize\catcode`\%=14\relax\special}%
  \setlength{\unitlength}{0.1bp}%
{\GNUPLOTspecial{!
/gnudict 256 dict def
gnudict begin
/Color false def
/Solid false def
/gnulinewidth 5.000 def
/userlinewidth gnulinewidth def
/vshift -33 def
/dl {10 mul} def
/hpt_ 31.5 def
/vpt_ 31.5 def
/hpt hpt_ def
/vpt vpt_ def
/M {moveto} bind def
/L {lineto} bind def
/R {rmoveto} bind def
/V {rlineto} bind def
/vpt2 vpt 2 mul def
/hpt2 hpt 2 mul def
/Lshow { currentpoint stroke M
  0 vshift R show } def
/Rshow { currentpoint stroke M
  dup stringwidth pop neg vshift R show } def
/Cshow { currentpoint stroke M
  dup stringwidth pop -2 div vshift R show } def
/UP { dup vpt_ mul /vpt exch def hpt_ mul /hpt exch def
  /hpt2 hpt 2 mul def /vpt2 vpt 2 mul def } def
/DL { Color {setrgbcolor Solid {pop []} if 0 setdash }
 {pop pop pop Solid {pop []} if 0 setdash} ifelse } def
/BL { stroke userlinewidth 2 mul setlinewidth } def
/AL { stroke userlinewidth 2 div setlinewidth } def
/UL { dup gnulinewidth mul /userlinewidth exch def
      10 mul /udl exch def } def
/PL { stroke userlinewidth setlinewidth } def
/LTb { BL [] 0 0 0 DL } def
/LTa { AL [1 udl mul 2 udl mul] 0 setdash 0 0 0 setrgbcolor } def
/LT0 { PL [] 1 0 0 DL } def
/LT1 { PL [4 dl 2 dl] 0 1 0 DL } def
/LT2 { PL [2 dl 3 dl] 0 0 1 DL } def
/LT3 { PL [1 dl 1.5 dl] 1 0 1 DL } def
/LT4 { PL [5 dl 2 dl 1 dl 2 dl] 0 1 1 DL } def
/LT5 { PL [4 dl 3 dl 1 dl 3 dl] 1 1 0 DL } def
/LT6 { PL [2 dl 2 dl 2 dl 4 dl] 0 0 0 DL } def
/LT7 { PL [2 dl 2 dl 2 dl 2 dl 2 dl 4 dl] 1 0.3 0 DL } def
/LT8 { PL [2 dl 2 dl 2 dl 2 dl 2 dl 2 dl 2 dl 4 dl] 0.5 0.5 0.5 DL } def
/Pnt { stroke [] 0 setdash
   gsave 1 setlinecap M 0 0 V stroke grestore } def
/Dia { stroke [] 0 setdash 2 copy vpt add M
  hpt neg vpt neg V hpt vpt neg V
  hpt vpt V hpt neg vpt V closepath stroke
  Pnt } def
/Pls { stroke [] 0 setdash vpt sub M 0 vpt2 V
  currentpoint stroke M
  hpt neg vpt neg R hpt2 0 V stroke
  } def
/Box { stroke [] 0 setdash 2 copy exch hpt sub exch vpt add M
  0 vpt2 neg V hpt2 0 V 0 vpt2 V
  hpt2 neg 0 V closepath stroke
  Pnt } def
/Crs { stroke [] 0 setdash exch hpt sub exch vpt add M
  hpt2 vpt2 neg V currentpoint stroke M
  hpt2 neg 0 R hpt2 vpt2 V stroke } def
/TriU { stroke [] 0 setdash 2 copy vpt 1.12 mul add M
  hpt neg vpt -1.62 mul V
  hpt 2 mul 0 V
  hpt neg vpt 1.62 mul V closepath stroke
  Pnt  } def
/Star { 2 copy Pls Crs } def
/BoxF { stroke [] 0 setdash exch hpt sub exch vpt add M
  0 vpt2 neg V  hpt2 0 V  0 vpt2 V
  hpt2 neg 0 V  closepath fill } def
/TriUF { stroke [] 0 setdash vpt 1.12 mul add M
  hpt neg vpt -1.62 mul V
  hpt 2 mul 0 V
  hpt neg vpt 1.62 mul V closepath fill } def
/TriD { stroke [] 0 setdash 2 copy vpt 1.12 mul sub M
  hpt neg vpt 1.62 mul V
  hpt 2 mul 0 V
  hpt neg vpt -1.62 mul V closepath stroke
  Pnt  } def
/TriDF { stroke [] 0 setdash vpt 1.12 mul sub M
  hpt neg vpt 1.62 mul V
  hpt 2 mul 0 V
  hpt neg vpt -1.62 mul V closepath fill} def
/DiaF { stroke [] 0 setdash vpt add M
  hpt neg vpt neg V hpt vpt neg V
  hpt vpt V hpt neg vpt V closepath fill } def
/Pent { stroke [] 0 setdash 2 copy gsave
  translate 0 hpt M 4 {72 rotate 0 hpt L} repeat
  closepath stroke grestore Pnt } def
/PentF { stroke [] 0 setdash gsave
  translate 0 hpt M 4 {72 rotate 0 hpt L} repeat
  closepath fill grestore } def
/Circle { stroke [] 0 setdash 2 copy
  hpt 0 360 arc stroke Pnt } def
/CircleF { stroke [] 0 setdash hpt 0 360 arc fill } def
/C0 { BL [] 0 setdash 2 copy moveto vpt 90 450  arc } bind def
/C1 { BL [] 0 setdash 2 copy        moveto
       2 copy  vpt 0 90 arc closepath fill
               vpt 0 360 arc closepath } bind def
/C2 { BL [] 0 setdash 2 copy moveto
       2 copy  vpt 90 180 arc closepath fill
               vpt 0 360 arc closepath } bind def
/C3 { BL [] 0 setdash 2 copy moveto
       2 copy  vpt 0 180 arc closepath fill
               vpt 0 360 arc closepath } bind def
/C4 { BL [] 0 setdash 2 copy moveto
       2 copy  vpt 180 270 arc closepath fill
               vpt 0 360 arc closepath } bind def
/C5 { BL [] 0 setdash 2 copy moveto
       2 copy  vpt 0 90 arc
       2 copy moveto
       2 copy  vpt 180 270 arc closepath fill
               vpt 0 360 arc } bind def
/C6 { BL [] 0 setdash 2 copy moveto
      2 copy  vpt 90 270 arc closepath fill
              vpt 0 360 arc closepath } bind def
/C7 { BL [] 0 setdash 2 copy moveto
      2 copy  vpt 0 270 arc closepath fill
              vpt 0 360 arc closepath } bind def
/C8 { BL [] 0 setdash 2 copy moveto
      2 copy vpt 270 360 arc closepath fill
              vpt 0 360 arc closepath } bind def
/C9 { BL [] 0 setdash 2 copy moveto
      2 copy  vpt 270 450 arc closepath fill
              vpt 0 360 arc closepath } bind def
/C10 { BL [] 0 setdash 2 copy 2 copy moveto vpt 270 360 arc closepath fill
       2 copy moveto
       2 copy vpt 90 180 arc closepath fill
               vpt 0 360 arc closepath } bind def
/C11 { BL [] 0 setdash 2 copy moveto
       2 copy  vpt 0 180 arc closepath fill
       2 copy moveto
       2 copy  vpt 270 360 arc closepath fill
               vpt 0 360 arc closepath } bind def
/C12 { BL [] 0 setdash 2 copy moveto
       2 copy  vpt 180 360 arc closepath fill
               vpt 0 360 arc closepath } bind def
/C13 { BL [] 0 setdash  2 copy moveto
       2 copy  vpt 0 90 arc closepath fill
       2 copy moveto
       2 copy  vpt 180 360 arc closepath fill
               vpt 0 360 arc closepath } bind def
/C14 { BL [] 0 setdash 2 copy moveto
       2 copy  vpt 90 360 arc closepath fill
               vpt 0 360 arc } bind def
/C15 { BL [] 0 setdash 2 copy vpt 0 360 arc closepath fill
               vpt 0 360 arc closepath } bind def
/Rec   { newpath 4 2 roll moveto 1 index 0 rlineto 0 exch rlineto
       neg 0 rlineto closepath } bind def
/Square { dup Rec } bind def
/Bsquare { vpt sub exch vpt sub exch vpt2 Square } bind def
/S0 { BL [] 0 setdash 2 copy moveto 0 vpt rlineto BL Bsquare } bind def
/S1 { BL [] 0 setdash 2 copy vpt Square fill Bsquare } bind def
/S2 { BL [] 0 setdash 2 copy exch vpt sub exch vpt Square fill Bsquare } bind def
/S3 { BL [] 0 setdash 2 copy exch vpt sub exch vpt2 vpt Rec fill Bsquare } bind def
/S4 { BL [] 0 setdash 2 copy exch vpt sub exch vpt sub vpt Square fill Bsquare } bind def
/S5 { BL [] 0 setdash 2 copy 2 copy vpt Square fill
       exch vpt sub exch vpt sub vpt Square fill Bsquare } bind def
/S6 { BL [] 0 setdash 2 copy exch vpt sub exch vpt sub vpt vpt2 Rec fill Bsquare } bind def
/S7 { BL [] 0 setdash 2 copy exch vpt sub exch vpt sub vpt vpt2 Rec fill
       2 copy vpt Square fill
       Bsquare } bind def
/S8 { BL [] 0 setdash 2 copy vpt sub vpt Square fill Bsquare } bind def
/S9 { BL [] 0 setdash 2 copy vpt sub vpt vpt2 Rec fill Bsquare } bind def
/S10 { BL [] 0 setdash 2 copy vpt sub vpt Square fill 2 copy exch vpt sub exch vpt Square fill
       Bsquare } bind def
/S11 { BL [] 0 setdash 2 copy vpt sub vpt Square fill 2 copy exch vpt sub exch vpt2 vpt Rec fill
       Bsquare } bind def
/S12 { BL [] 0 setdash 2 copy exch vpt sub exch vpt sub vpt2 vpt Rec fill Bsquare } bind def
/S13 { BL [] 0 setdash 2 copy exch vpt sub exch vpt sub vpt2 vpt Rec fill
       2 copy vpt Square fill Bsquare } bind def
/S14 { BL [] 0 setdash 2 copy exch vpt sub exch vpt sub vpt2 vpt Rec fill
       2 copy exch vpt sub exch vpt Square fill Bsquare } bind def
/S15 { BL [] 0 setdash 2 copy Bsquare fill Bsquare } bind def
/D0 { gsave translate 45 rotate 0 0 S0 stroke grestore } bind def
/D1 { gsave translate 45 rotate 0 0 S1 stroke grestore } bind def
/D2 { gsave translate 45 rotate 0 0 S2 stroke grestore } bind def
/D3 { gsave translate 45 rotate 0 0 S3 stroke grestore } bind def
/D4 { gsave translate 45 rotate 0 0 S4 stroke grestore } bind def
/D5 { gsave translate 45 rotate 0 0 S5 stroke grestore } bind def
/D6 { gsave translate 45 rotate 0 0 S6 stroke grestore } bind def
/D7 { gsave translate 45 rotate 0 0 S7 stroke grestore } bind def
/D8 { gsave translate 45 rotate 0 0 S8 stroke grestore } bind def
/D9 { gsave translate 45 rotate 0 0 S9 stroke grestore } bind def
/D10 { gsave translate 45 rotate 0 0 S10 stroke grestore } bind def
/D11 { gsave translate 45 rotate 0 0 S11 stroke grestore } bind def
/D12 { gsave translate 45 rotate 0 0 S12 stroke grestore } bind def
/D13 { gsave translate 45 rotate 0 0 S13 stroke grestore } bind def
/D14 { gsave translate 45 rotate 0 0 S14 stroke grestore } bind def
/D15 { gsave translate 45 rotate 0 0 S15 stroke grestore } bind def
/DiaE { stroke [] 0 setdash vpt add M
  hpt neg vpt neg V hpt vpt neg V
  hpt vpt V hpt neg vpt V closepath stroke } def
/BoxE { stroke [] 0 setdash exch hpt sub exch vpt add M
  0 vpt2 neg V hpt2 0 V 0 vpt2 V
  hpt2 neg 0 V closepath stroke } def
/TriUE { stroke [] 0 setdash vpt 1.12 mul add M
  hpt neg vpt -1.62 mul V
  hpt 2 mul 0 V
  hpt neg vpt 1.62 mul V closepath stroke } def
/TriDE { stroke [] 0 setdash vpt 1.12 mul sub M
  hpt neg vpt 1.62 mul V
  hpt 2 mul 0 V
  hpt neg vpt -1.62 mul V closepath stroke } def
/PentE { stroke [] 0 setdash gsave
  translate 0 hpt M 4 {72 rotate 0 hpt L} repeat
  closepath stroke grestore } def
/CircE { stroke [] 0 setdash 
  hpt 0 360 arc stroke } def
/Opaque { gsave closepath 1 setgray fill grestore 0 setgray closepath } def
/DiaW { stroke [] 0 setdash vpt add M
  hpt neg vpt neg V hpt vpt neg V
  hpt vpt V hpt neg vpt V Opaque stroke } def
/BoxW { stroke [] 0 setdash exch hpt sub exch vpt add M
  0 vpt2 neg V hpt2 0 V 0 vpt2 V
  hpt2 neg 0 V Opaque stroke } def
/TriUW { stroke [] 0 setdash vpt 1.12 mul add M
  hpt neg vpt -1.62 mul V
  hpt 2 mul 0 V
  hpt neg vpt 1.62 mul V Opaque stroke } def
/TriDW { stroke [] 0 setdash vpt 1.12 mul sub M
  hpt neg vpt 1.62 mul V
  hpt 2 mul 0 V
  hpt neg vpt -1.62 mul V Opaque stroke } def
/PentW { stroke [] 0 setdash gsave
  translate 0 hpt M 4 {72 rotate 0 hpt L} repeat
  Opaque stroke grestore } def
/CircW { stroke [] 0 setdash 
  hpt 0 360 arc Opaque stroke } def
/BoxFill { gsave Rec 1 setgray fill grestore } def
end
}}%
\begin{picture}(3600,2160)(0,-100)%
{\GNUPLOTspecial{"
gnudict begin
gsave
0 0 translate
0.100 0.100 scale
0 setgray
newpath
1.000 UL
LTb
200 300 M
0 18 V
0 1742 R
0 -18 V
606 300 M
0 18 V
0 1742 R
0 -18 V
1013 300 M
0 18 V
0 1742 R
0 -18 V
1419 300 M
0 18 V
0 1742 R
0 -18 V
1825 300 M
0 18 V
0 1742 R
0 -18 V
2231 300 M
0 18 V
0 1742 R
0 -18 V
2638 300 M
0 18 V
0 1742 R
0 -18 V
3044 300 M
0 18 V
0 1742 R
0 -18 V
3450 300 M
0 18 V
0 1742 R
0 -18 V
1.000 UL
LTb
200 300 M
3250 0 V
0 1760 V
-3250 0 V
200 300 L
1.000 UL
LT0
200 302 M
8 0 V
22 -1 V
21 0 V
22 -1 V
21 0 V
22 0 V
21 0 V
22 0 V
21 0 V
22 0 V
22 0 V
21 0 V
22 0 V
21 0 V
22 0 V
21 0 V
22 0 V
22 0 V
21 0 V
22 0 V
21 0 V
22 0 V
21 0 V
22 0 V
21 0 V
22 0 V
22 0 V
21 0 V
22 0 V
21 0 V
22 0 V
21 1 V
22 0 V
21 1 V
22 2 V
22 8 V
21 42 V
22 54 V
21 -13 V
22 -59 V
21 -29 V
22 -3 V
22 -3 V
21 0 V
22 -1 V
21 0 V
22 0 V
21 0 V
22 0 V
21 0 V
22 0 V
22 0 V
21 0 V
22 0 V
21 0 V
22 0 V
21 0 V
22 0 V
21 0 V
22 0 V
22 0 V
21 0 V
22 0 V
21 0 V
22 0 V
21 1 V
22 0 V
22 1 V
21 0 V
22 2 V
21 1 V
22 12 V
21 6 V
22 178 V
21 821 V
22 625 V
22 -635 V
21 -815 V
22 -174 V
21 -6 V
22 -12 V
21 -1 V
22 -2 V
21 0 V
22 -1 V
22 0 V
21 -1 V
22 0 V
21 0 V
22 0 V
21 0 V
22 0 V
22 0 V
21 0 V
22 0 V
21 0 V
22 0 V
21 0 V
22 0 V
21 0 V
22 0 V
22 0 V
21 0 V
22 0 V
21 0 V
22 0 V
21 1 V
22 0 V
21 1 V
22 4 V
22 6 V
21 46 V
22 101 V
21 26 V
22 -93 V
21 -71 V
22 -12 V
22 -5 V
21 -2 V
22 -1 V
21 0 V
22 -1 V
21 0 V
22 0 V
21 0 V
22 0 V
22 0 V
21 0 V
22 0 V
21 0 V
22 0 V
21 0 V
22 0 V
21 0 V
22 0 V
22 0 V
21 0 V
22 0 V
21 0 V
22 0 V
21 0 V
22 0 V
22 0 V
21 0 V
22 0 V
21 0 V
22 0 V
21 1 V
22 0 V
21 2 V
22 2 V
8 0 V
stroke
grestore
end
showpage
}}%
\put(1825,50){\makebox(0,0){$p_z \left[ \hbar/a_\perp \right]$}}%
\put(100,1180){%
\special{ps: gsave currentpoint currentpoint translate
270 rotate neg exch neg exch translate}%
\makebox(0,0)[b]{\shortstack{$Q \ S(Q,\Omega)/m$}}%
\special{ps: currentpoint grestore moveto}%
}%
\put(3450,200){\makebox(0,0){$2$}}%
\put(3044,200){\makebox(0,0){$1.5$}}%
\put(2638,200){\makebox(0,0){$1$}}%
\put(2231,200){\makebox(0,0){$0.5$}}%
\put(1825,200){\makebox(0,0){$0$}}%
\put(1419,200){\makebox(0,0){$-0.5$}}%
\put(1013,200){\makebox(0,0){$-1$}}%
\put(606,200){\makebox(0,0){$-1.5$}}%
\put(200,200){\makebox(0,0){$-2$}}%
\end{picture}%
\endgroup
 

%% file: fig3.tex
\begingroup%
  \makeatletter%
  \newcommand{\GNUPLOTspecial}{%
    \@sanitize\catcode`\%=14\relax\special}%
  \setlength{\unitlength}{0.1bp}%
{\GNUPLOTspecial{!
/gnudict 256 dict def
gnudict begin
/Color false def
/Solid false def
/gnulinewidth 5.000 def
/userlinewidth gnulinewidth def
/vshift -33 def
/dl {10 mul} def
/hpt_ 31.5 def
/vpt_ 31.5 def
/hpt hpt_ def
/vpt vpt_ def
/M {moveto} bind def
/L {lineto} bind def
/R {rmoveto} bind def
/V {rlineto} bind def
/vpt2 vpt 2 mul def
/hpt2 hpt 2 mul def
/Lshow { currentpoint stroke M
  0 vshift R show } def
/Rshow { currentpoint stroke M
  dup stringwidth pop neg vshift R show } def
/Cshow { currentpoint stroke M
  dup stringwidth pop -2 div vshift R show } def
/UP { dup vpt_ mul /vpt exch def hpt_ mul /hpt exch def
  /hpt2 hpt 2 mul def /vpt2 vpt 2 mul def } def
/DL { Color {setrgbcolor Solid {pop []} if 0 setdash }
 {pop pop pop Solid {pop []} if 0 setdash} ifelse } def
/BL { stroke userlinewidth 2 mul setlinewidth } def
/AL { stroke userlinewidth 2 div setlinewidth } def
/UL { dup gnulinewidth mul /userlinewidth exch def
      10 mul /udl exch def } def
/PL { stroke userlinewidth setlinewidth } def
/LTb { BL [] 0 0 0 DL } def
/LTa { AL [1 udl mul 2 udl mul] 0 setdash 0 0 0 setrgbcolor } def
/LT0 { PL [] 1 0 0 DL } def
/LT1 { PL [4 dl 2 dl] 0 1 0 DL } def
/LT2 { PL [2 dl 3 dl] 0 0 1 DL } def
/LT3 { PL [1 dl 1.5 dl] 1 0 1 DL } def
/LT4 { PL [5 dl 2 dl 1 dl 2 dl] 0 1 1 DL } def
/LT5 { PL [4 dl 3 dl 1 dl 3 dl] 1 1 0 DL } def
/LT6 { PL [2 dl 2 dl 2 dl 4 dl] 0 0 0 DL } def
/LT7 { PL [2 dl 2 dl 2 dl 2 dl 2 dl 4 dl] 1 0.3 0 DL } def
/LT8 { PL [2 dl 2 dl 2 dl 2 dl 2 dl 2 dl 2 dl 4 dl] 0.5 0.5 0.5 DL } def
/Pnt { stroke [] 0 setdash
   gsave 1 setlinecap M 0 0 V stroke grestore } def
/Dia { stroke [] 0 setdash 2 copy vpt add M
  hpt neg vpt neg V hpt vpt neg V
  hpt vpt V hpt neg vpt V closepath stroke
  Pnt } def
/Pls { stroke [] 0 setdash vpt sub M 0 vpt2 V
  currentpoint stroke M
  hpt neg vpt neg R hpt2 0 V stroke
  } def
/Box { stroke [] 0 setdash 2 copy exch hpt sub exch vpt add M
  0 vpt2 neg V hpt2 0 V 0 vpt2 V
  hpt2 neg 0 V closepath stroke
  Pnt } def
/Crs { stroke [] 0 setdash exch hpt sub exch vpt add M
  hpt2 vpt2 neg V currentpoint stroke M
  hpt2 neg 0 R hpt2 vpt2 V stroke } def
/TriU { stroke [] 0 setdash 2 copy vpt 1.12 mul add M
  hpt neg vpt -1.62 mul V
  hpt 2 mul 0 V
  hpt neg vpt 1.62 mul V closepath stroke
  Pnt  } def
/Star { 2 copy Pls Crs } def
/BoxF { stroke [] 0 setdash exch hpt sub exch vpt add M
  0 vpt2 neg V  hpt2 0 V  0 vpt2 V
  hpt2 neg 0 V  closepath fill } def
/TriUF { stroke [] 0 setdash vpt 1.12 mul add M
  hpt neg vpt -1.62 mul V
  hpt 2 mul 0 V
  hpt neg vpt 1.62 mul V closepath fill } def
/TriD { stroke [] 0 setdash 2 copy vpt 1.12 mul sub M
  hpt neg vpt 1.62 mul V
  hpt 2 mul 0 V
  hpt neg vpt -1.62 mul V closepath stroke
  Pnt  } def
/TriDF { stroke [] 0 setdash vpt 1.12 mul sub M
  hpt neg vpt 1.62 mul V
  hpt 2 mul 0 V
  hpt neg vpt -1.62 mul V closepath fill} def
/DiaF { stroke [] 0 setdash vpt add M
  hpt neg vpt neg V hpt vpt neg V
  hpt vpt V hpt neg vpt V closepath fill } def
/Pent { stroke [] 0 setdash 2 copy gsave
  translate 0 hpt M 4 {72 rotate 0 hpt L} repeat
  closepath stroke grestore Pnt } def
/PentF { stroke [] 0 setdash gsave
  translate 0 hpt M 4 {72 rotate 0 hpt L} repeat
  closepath fill grestore } def
/Circle { stroke [] 0 setdash 2 copy
  hpt 0 360 arc stroke Pnt } def
/CircleF { stroke [] 0 setdash hpt 0 360 arc fill } def
/C0 { BL [] 0 setdash 2 copy moveto vpt 90 450  arc } bind def
/C1 { BL [] 0 setdash 2 copy        moveto
       2 copy  vpt 0 90 arc closepath fill
               vpt 0 360 arc closepath } bind def
/C2 { BL [] 0 setdash 2 copy moveto
       2 copy  vpt 90 180 arc closepath fill
               vpt 0 360 arc closepath } bind def
/C3 { BL [] 0 setdash 2 copy moveto
       2 copy  vpt 0 180 arc closepath fill
               vpt 0 360 arc closepath } bind def
/C4 { BL [] 0 setdash 2 copy moveto
       2 copy  vpt 180 270 arc closepath fill
               vpt 0 360 arc closepath } bind def
/C5 { BL [] 0 setdash 2 copy moveto
       2 copy  vpt 0 90 arc
       2 copy moveto
       2 copy  vpt 180 270 arc closepath fill
               vpt 0 360 arc } bind def
/C6 { BL [] 0 setdash 2 copy moveto
      2 copy  vpt 90 270 arc closepath fill
              vpt 0 360 arc closepath } bind def
/C7 { BL [] 0 setdash 2 copy moveto
      2 copy  vpt 0 270 arc closepath fill
              vpt 0 360 arc closepath } bind def
/C8 { BL [] 0 setdash 2 copy moveto
      2 copy vpt 270 360 arc closepath fill
              vpt 0 360 arc closepath } bind def
/C9 { BL [] 0 setdash 2 copy moveto
      2 copy  vpt 270 450 arc closepath fill
              vpt 0 360 arc closepath } bind def
/C10 { BL [] 0 setdash 2 copy 2 copy moveto vpt 270 360 arc closepath fill
       2 copy moveto
       2 copy vpt 90 180 arc closepath fill
               vpt 0 360 arc closepath } bind def
/C11 { BL [] 0 setdash 2 copy moveto
       2 copy  vpt 0 180 arc closepath fill
       2 copy moveto
       2 copy  vpt 270 360 arc closepath fill
               vpt 0 360 arc closepath } bind def
/C12 { BL [] 0 setdash 2 copy moveto
       2 copy  vpt 180 360 arc closepath fill
               vpt 0 360 arc closepath } bind def
/C13 { BL [] 0 setdash  2 copy moveto
       2 copy  vpt 0 90 arc closepath fill
       2 copy moveto
       2 copy  vpt 180 360 arc closepath fill
               vpt 0 360 arc closepath } bind def
/C14 { BL [] 0 setdash 2 copy moveto
       2 copy  vpt 90 360 arc closepath fill
               vpt 0 360 arc } bind def
/C15 { BL [] 0 setdash 2 copy vpt 0 360 arc closepath fill
               vpt 0 360 arc closepath } bind def
/Rec   { newpath 4 2 roll moveto 1 index 0 rlineto 0 exch rlineto
       neg 0 rlineto closepath } bind def
/Square { dup Rec } bind def
/Bsquare { vpt sub exch vpt sub exch vpt2 Square } bind def
/S0 { BL [] 0 setdash 2 copy moveto 0 vpt rlineto BL Bsquare } bind def
/S1 { BL [] 0 setdash 2 copy vpt Square fill Bsquare } bind def
/S2 { BL [] 0 setdash 2 copy exch vpt sub exch vpt Square fill Bsquare } bind def
/S3 { BL [] 0 setdash 2 copy exch vpt sub exch vpt2 vpt Rec fill Bsquare } bind def
/S4 { BL [] 0 setdash 2 copy exch vpt sub exch vpt sub vpt Square fill Bsquare } bind def
/S5 { BL [] 0 setdash 2 copy 2 copy vpt Square fill
       exch vpt sub exch vpt sub vpt Square fill Bsquare } bind def
/S6 { BL [] 0 setdash 2 copy exch vpt sub exch vpt sub vpt vpt2 Rec fill Bsquare } bind def
/S7 { BL [] 0 setdash 2 copy exch vpt sub exch vpt sub vpt vpt2 Rec fill
       2 copy vpt Square fill
       Bsquare } bind def
/S8 { BL [] 0 setdash 2 copy vpt sub vpt Square fill Bsquare } bind def
/S9 { BL [] 0 setdash 2 copy vpt sub vpt vpt2 Rec fill Bsquare } bind def
/S10 { BL [] 0 setdash 2 copy vpt sub vpt Square fill 2 copy exch vpt sub exch vpt Square fill
       Bsquare } bind def
/S11 { BL [] 0 setdash 2 copy vpt sub vpt Square fill 2 copy exch vpt sub exch vpt2 vpt Rec fill
       Bsquare } bind def
/S12 { BL [] 0 setdash 2 copy exch vpt sub exch vpt sub vpt2 vpt Rec fill Bsquare } bind def
/S13 { BL [] 0 setdash 2 copy exch vpt sub exch vpt sub vpt2 vpt Rec fill
       2 copy vpt Square fill Bsquare } bind def
/S14 { BL [] 0 setdash 2 copy exch vpt sub exch vpt sub vpt2 vpt Rec fill
       2 copy exch vpt sub exch vpt Square fill Bsquare } bind def
/S15 { BL [] 0 setdash 2 copy Bsquare fill Bsquare } bind def
/D0 { gsave translate 45 rotate 0 0 S0 stroke grestore } bind def
/D1 { gsave translate 45 rotate 0 0 S1 stroke grestore } bind def
/D2 { gsave translate 45 rotate 0 0 S2 stroke grestore } bind def
/D3 { gsave translate 45 rotate 0 0 S3 stroke grestore } bind def
/D4 { gsave translate 45 rotate 0 0 S4 stroke grestore } bind def
/D5 { gsave translate 45 rotate 0 0 S5 stroke grestore } bind def
/D6 { gsave translate 45 rotate 0 0 S6 stroke grestore } bind def
/D7 { gsave translate 45 rotate 0 0 S7 stroke grestore } bind def
/D8 { gsave translate 45 rotate 0 0 S8 stroke grestore } bind def
/D9 { gsave translate 45 rotate 0 0 S9 stroke grestore } bind def
/D10 { gsave translate 45 rotate 0 0 S10 stroke grestore } bind def
/D11 { gsave translate 45 rotate 0 0 S11 stroke grestore } bind def
/D12 { gsave translate 45 rotate 0 0 S12 stroke grestore } bind def
/D13 { gsave translate 45 rotate 0 0 S13 stroke grestore } bind def
/D14 { gsave translate 45 rotate 0 0 S14 stroke grestore } bind def
/D15 { gsave translate 45 rotate 0 0 S15 stroke grestore } bind def
/DiaE { stroke [] 0 setdash vpt add M
  hpt neg vpt neg V hpt vpt neg V
  hpt vpt V hpt neg vpt V closepath stroke } def
/BoxE { stroke [] 0 setdash exch hpt sub exch vpt add M
  0 vpt2 neg V hpt2 0 V 0 vpt2 V
  hpt2 neg 0 V closepath stroke } def
/TriUE { stroke [] 0 setdash vpt 1.12 mul add M
  hpt neg vpt -1.62 mul V
  hpt 2 mul 0 V
  hpt neg vpt 1.62 mul V closepath stroke } def
/TriDE { stroke [] 0 setdash vpt 1.12 mul sub M
  hpt neg vpt 1.62 mul V
  hpt 2 mul 0 V
  hpt neg vpt -1.62 mul V closepath stroke } def
/PentE { stroke [] 0 setdash gsave
  translate 0 hpt M 4 {72 rotate 0 hpt L} repeat
  closepath stroke grestore } def
/CircE { stroke [] 0 setdash 
  hpt 0 360 arc stroke } def
/Opaque { gsave closepath 1 setgray fill grestore 0 setgray closepath } def
/DiaW { stroke [] 0 setdash vpt add M
  hpt neg vpt neg V hpt vpt neg V
  hpt vpt V hpt neg vpt V Opaque stroke } def
/BoxW { stroke [] 0 setdash exch hpt sub exch vpt add M
  0 vpt2 neg V hpt2 0 V 0 vpt2 V
  hpt2 neg 0 V Opaque stroke } def
/TriUW { stroke [] 0 setdash vpt 1.12 mul add M
  hpt neg vpt -1.62 mul V
  hpt 2 mul 0 V
  hpt neg vpt 1.62 mul V Opaque stroke } def
/TriDW { stroke [] 0 setdash vpt 1.12 mul sub M
  hpt neg vpt 1.62 mul V
  hpt 2 mul 0 V
  hpt neg vpt -1.62 mul V Opaque stroke } def
/PentW { stroke [] 0 setdash gsave
  translate 0 hpt M 4 {72 rotate 0 hpt L} repeat
  Opaque stroke grestore } def
/CircW { stroke [] 0 setdash 
  hpt 0 360 arc Opaque stroke } def
/BoxFill { gsave Rec 1 setgray fill grestore } def
end
}}%
\begin{picture}(3600,2160)(0,-100)%
{\GNUPLOTspecial{"
gnudict begin
gsave
0 0 translate
0.100 0.100 scale
0 setgray
newpath
1.000 UL
LTb
500 300 M
18 0 V
2932 0 R
-18 0 V
500 507 M
18 0 V
2932 0 R
-18 0 V
500 714 M
18 0 V
2932 0 R
-18 0 V
500 921 M
18 0 V
2932 0 R
-18 0 V
500 1128 M
18 0 V
2932 0 R
-18 0 V
500 1335 M
18 0 V
2932 0 R
-18 0 V
500 1542 M
18 0 V
2932 0 R
-18 0 V
500 1749 M
18 0 V
2932 0 R
-18 0 V
500 1956 M
18 0 V
2932 0 R
-18 0 V
500 300 M
0 18 V
0 1742 R
0 -18 V
961 300 M
0 18 V
0 1742 R
0 -18 V
1422 300 M
0 18 V
0 1742 R
0 -18 V
1883 300 M
0 18 V
0 1742 R
0 -18 V
2344 300 M
0 18 V
0 1742 R
0 -18 V
2805 300 M
0 18 V
0 1742 R
0 -18 V
3266 300 M
0 18 V
0 1742 R
0 -18 V
1.000 UL
LTb
500 300 M
2950 0 V
0 1760 V
-2950 0 V
500 300 L
1.000 UL
LT0
1265 2060 M
15 -91 V
22 -113 V
22 -94 V
22 -76 V
22 -62 V
22 -48 V
22 -37 V
22 -26 V
22 -16 V
22 -8 V
22 -2 V
22 5 V
22 9 V
22 14 V
22 17 V
22 19 V
22 22 V
22 22 V
22 23 V
22 24 V
22 22 V
22 22 V
22 21 V
22 19 V
22 17 V
22 15 V
22 13 V
22 10 V
22 8 V
22 6 V
22 3 V
22 0 V
22 -2 V
22 -4 V
22 -6 V
22 -8 V
22 -10 V
22 -11 V
22 -12 V
22 -12 V
22 -13 V
22 -14 V
22 -14 V
22 -13 V
22 -13 V
22 -12 V
22 -12 V
22 -10 V
22 -10 V
22 -9 V
22 -8 V
22 -7 V
22 -6 V
22 -5 V
22 -4 V
22 -3 V
22 -2 V
22 -2 V
22 -2 V
22 -1 V
22 -2 V
22 -1 V
22 -1 V
22 -1 V
22 -2 V
22 -2 V
22 -2 V
22 -3 V
22 -4 V
22 -4 V
22 -6 V
22 -6 V
22 -6 V
22 -8 V
22 -8 V
22 -8 V
22 -10 V
22 -9 V
22 -11 V
22 -10 V
22 -11 V
22 -10 V
22 -11 V
22 -10 V
22 -10 V
22 -10 V
22 -9 V
22 -9 V
22 -7 V
22 -7 V
22 -6 V
22 -5 V
22 -4 V
22 -3 V
22 -1 V
22 -1 V
22 1 V
22 2 V
22 2 V
14 3 V
1.000 UL
LT0
919 2060 M
8 -99 V
22 -219 V
22 -180 V
22 -147 V
22 -117 V
23 -92 V
22 -69 V
22 -49 V
22 -34 V
22 -19 V
22 -7 V
22 2 V
22 9 V
22 15 V
22 19 V
22 22 V
22 23 V
22 23 V
22 22 V
22 20 V
22 17 V
22 15 V
22 11 V
22 7 V
22 3 V
22 0 V
22 -4 V
22 -7 V
22 -9 V
22 -11 V
22 -13 V
22 -14 V
22 -13 V
22 -14 V
22 -13 V
22 -12 V
22 -10 V
22 -9 V
22 -7 V
22 -5 V
22 -3 V
22 -2 V
22 -1 V
22 1 V
22 2 V
22 2 V
22 3 V
22 3 V
22 2 V
22 3 V
22 1 V
22 1 V
22 -1 V
22 -1 V
22 -2 V
22 -4 V
22 -4 V
22 -6 V
22 -6 V
22 -6 V
22 -7 V
22 -7 V
22 -6 V
22 -7 V
22 -6 V
22 -5 V
22 -4 V
22 -4 V
22 -2 V
22 -1 V
22 0 V
22 0 V
22 2 V
22 3 V
22 4 V
22 4 V
22 6 V
22 5 V
22 6 V
22 6 V
22 6 V
22 6 V
22 5 V
22 5 V
22 5 V
22 3 V
22 3 V
22 3 V
22 1 V
22 1 V
22 0 V
22 -1 V
22 -1 V
22 -2 V
22 -3 V
22 -3 V
22 -3 V
22 -3 V
22 -4 V
22 -4 V
22 -4 V
22 -3 V
22 -4 V
22 -3 V
22 -3 V
22 -3 V
22 -2 V
22 -2 V
22 -2 V
22 -1 V
22 -1 V
22 -1 V
22 0 V
22 0 V
22 0 V
14 0 V
1.000 UL
LT0
756 2060 M
17 -246 V
22 -251 V
22 -205 V
22 -164 V
22 -129 V
22 -98 V
22 -71 V
22 -49 V
22 -30 V
22 -15 V
22 -2 V
22 6 V
23 13 V
22 17 V
22 19 V
22 19 V
22 18 V
22 15 V
22 11 V
22 7 V
22 3 V
22 -1 V
22 -5 V
22 -7 V
22 -11 V
22 -11 V
22 -12 V
22 -12 V
22 -10 V
22 -10 V
22 -7 V
22 -6 V
22 -3 V
22 -2 V
22 0 V
22 2 V
22 2 V
22 2 V
22 3 V
22 2 V
22 1 V
22 1 V
22 -1 V
22 -2 V
22 -3 V
22 -3 V
22 -5 V
22 -5 V
22 -5 V
22 -5 V
22 -4 V
22 -4 V
22 -3 V
22 -2 V
22 -2 V
22 0 V
22 1 V
22 1 V
22 3 V
22 2 V
22 3 V
22 3 V
22 3 V
22 3 V
22 2 V
22 2 V
22 1 V
22 0 V
22 0 V
22 -1 V
22 -1 V
22 -2 V
22 -2 V
22 -3 V
22 -2 V
22 -3 V
22 -2 V
22 -2 V
22 -2 V
22 -2 V
22 -2 V
22 -2 V
22 -1 V
22 -2 V
22 -1 V
22 -2 V
22 -1 V
22 -2 V
22 -2 V
22 -2 V
22 -2 V
22 -2 V
22 -3 V
22 -2 V
22 -2 V
22 -2 V
22 -3 V
22 -2 V
22 -2 V
22 -2 V
22 -2 V
22 -2 V
22 -3 V
22 -2 V
22 -3 V
22 -3 V
22 -4 V
22 -4 V
22 -5 V
22 -5 V
22 -5 V
22 -6 V
22 -7 V
22 -6 V
22 -7 V
22 -7 V
22 -7 V
22 -6 V
22 -7 V
22 -6 V
22 -5 V
22 -5 V
14 -2 V
stroke
grestore
end
showpage
}}%
\put(1837,1284){\makebox(0,0)[l]{$v^2 = 0.93$}}%
\put(2344,1749){\makebox(0,0)[l]{$v^2 = 1.59$}}%
\put(1837,662){\makebox(0,0)[l]{$v^2 = 0.61$}}%
\put(1975,50){\makebox(0,0){$t_B \left[ 2 \pi / \omega_\perp \right]$}}%
\put(100,1180){%
\special{ps: gsave currentpoint currentpoint translate
270 rotate neg exch neg exch translate}%
\makebox(0,0)[b]{\shortstack{$v^2_q$}}%
\special{ps: currentpoint grestore moveto}%
}%
\put(3266,200){\makebox(0,0){$0.3$}}%
\put(2805,200){\makebox(0,0){$0.25$}}%
\put(2344,200){\makebox(0,0){$0.2$}}%
\put(1883,200){\makebox(0,0){$0.15$}}%
\put(1422,200){\makebox(0,0){$0.1$}}%
\put(961,200){\makebox(0,0){$0.05$}}%
\put(500,200){\makebox(0,0){$0$}}%
\put(450,1956){\makebox(0,0)[r]{$1.6$}}%
\put(450,1749){\makebox(0,0)[r]{$1.4$}}%
\put(450,1542){\makebox(0,0)[r]{$1.2$}}%
\put(450,1335){\makebox(0,0)[r]{$1$}}%
\put(450,1128){\makebox(0,0)[r]{$0.8$}}%
\put(450,921){\makebox(0,0)[r]{$0.6$}}%
\put(450,714){\makebox(0,0)[r]{$0.4$}}%
\put(450,507){\makebox(0,0)[r]{$0.2$}}%
\put(450,300){\makebox(0,0)[r]{$0$}}%
\end{picture}%
\endgroup
 

%% file: fig4.tex
\begingroup%
  \makeatletter%
  \newcommand{\GNUPLOTspecial}{%
    \@sanitize\catcode`\%=14\relax\special}%
  \setlength{\unitlength}{0.1bp}%
{\GNUPLOTspecial{!
/gnudict 256 dict def
gnudict begin
/Color false def
/Solid false def
/gnulinewidth 5.000 def
/userlinewidth gnulinewidth def
/vshift -33 def
/dl {10 mul} def
/hpt_ 31.5 def
/vpt_ 31.5 def
/hpt hpt_ def
/vpt vpt_ def
/M {moveto} bind def
/L {lineto} bind def
/R {rmoveto} bind def
/V {rlineto} bind def
/vpt2 vpt 2 mul def
/hpt2 hpt 2 mul def
/Lshow { currentpoint stroke M
  0 vshift R show } def
/Rshow { currentpoint stroke M
  dup stringwidth pop neg vshift R show } def
/Cshow { currentpoint stroke M
  dup stringwidth pop -2 div vshift R show } def
/UP { dup vpt_ mul /vpt exch def hpt_ mul /hpt exch def
  /hpt2 hpt 2 mul def /vpt2 vpt 2 mul def } def
/DL { Color {setrgbcolor Solid {pop []} if 0 setdash }
 {pop pop pop Solid {pop []} if 0 setdash} ifelse } def
/BL { stroke userlinewidth 2 mul setlinewidth } def
/AL { stroke userlinewidth 2 div setlinewidth } def
/UL { dup gnulinewidth mul /userlinewidth exch def
      10 mul /udl exch def } def
/PL { stroke userlinewidth setlinewidth } def
/LTb { BL [] 0 0 0 DL } def
/LTa { AL [1 udl mul 2 udl mul] 0 setdash 0 0 0 setrgbcolor } def
/LT0 { PL [] 1 0 0 DL } def
/LT1 { PL [4 dl 2 dl] 0 1 0 DL } def
/LT2 { PL [2 dl 3 dl] 0 0 1 DL } def
/LT3 { PL [1 dl 1.5 dl] 1 0 1 DL } def
/LT4 { PL [5 dl 2 dl 1 dl 2 dl] 0 1 1 DL } def
/LT5 { PL [4 dl 3 dl 1 dl 3 dl] 1 1 0 DL } def
/LT6 { PL [2 dl 2 dl 2 dl 4 dl] 0 0 0 DL } def
/LT7 { PL [2 dl 2 dl 2 dl 2 dl 2 dl 4 dl] 1 0.3 0 DL } def
/LT8 { PL [2 dl 2 dl 2 dl 2 dl 2 dl 2 dl 2 dl 4 dl] 0.5 0.5 0.5 DL } def
/Pnt { stroke [] 0 setdash
   gsave 1 setlinecap M 0 0 V stroke grestore } def
/Dia { stroke [] 0 setdash 2 copy vpt add M
  hpt neg vpt neg V hpt vpt neg V
  hpt vpt V hpt neg vpt V closepath stroke
  Pnt } def
/Pls { stroke [] 0 setdash vpt sub M 0 vpt2 V
  currentpoint stroke M
  hpt neg vpt neg R hpt2 0 V stroke
  } def
/Box { stroke [] 0 setdash 2 copy exch hpt sub exch vpt add M
  0 vpt2 neg V hpt2 0 V 0 vpt2 V
  hpt2 neg 0 V closepath stroke
  Pnt } def
/Crs { stroke [] 0 setdash exch hpt sub exch vpt add M
  hpt2 vpt2 neg V currentpoint stroke M
  hpt2 neg 0 R hpt2 vpt2 V stroke } def
/TriU { stroke [] 0 setdash 2 copy vpt 1.12 mul add M
  hpt neg vpt -1.62 mul V
  hpt 2 mul 0 V
  hpt neg vpt 1.62 mul V closepath stroke
  Pnt  } def
/Star { 2 copy Pls Crs } def
/BoxF { stroke [] 0 setdash exch hpt sub exch vpt add M
  0 vpt2 neg V  hpt2 0 V  0 vpt2 V
  hpt2 neg 0 V  closepath fill } def
/TriUF { stroke [] 0 setdash vpt 1.12 mul add M
  hpt neg vpt -1.62 mul V
  hpt 2 mul 0 V
  hpt neg vpt 1.62 mul V closepath fill } def
/TriD { stroke [] 0 setdash 2 copy vpt 1.12 mul sub M
  hpt neg vpt 1.62 mul V
  hpt 2 mul 0 V
  hpt neg vpt -1.62 mul V closepath stroke
  Pnt  } def
/TriDF { stroke [] 0 setdash vpt 1.12 mul sub M
  hpt neg vpt 1.62 mul V
  hpt 2 mul 0 V
  hpt neg vpt -1.62 mul V closepath fill} def
/DiaF { stroke [] 0 setdash vpt add M
  hpt neg vpt neg V hpt vpt neg V
  hpt vpt V hpt neg vpt V closepath fill } def
/Pent { stroke [] 0 setdash 2 copy gsave
  translate 0 hpt M 4 {72 rotate 0 hpt L} repeat
  closepath stroke grestore Pnt } def
/PentF { stroke [] 0 setdash gsave
  translate 0 hpt M 4 {72 rotate 0 hpt L} repeat
  closepath fill grestore } def
/Circle { stroke [] 0 setdash 2 copy
  hpt 0 360 arc stroke Pnt } def
/CircleF { stroke [] 0 setdash hpt 0 360 arc fill } def
/C0 { BL [] 0 setdash 2 copy moveto vpt 90 450  arc } bind def
/C1 { BL [] 0 setdash 2 copy        moveto
       2 copy  vpt 0 90 arc closepath fill
               vpt 0 360 arc closepath } bind def
/C2 { BL [] 0 setdash 2 copy moveto
       2 copy  vpt 90 180 arc closepath fill
               vpt 0 360 arc closepath } bind def
/C3 { BL [] 0 setdash 2 copy moveto
       2 copy  vpt 0 180 arc closepath fill
               vpt 0 360 arc closepath } bind def
/C4 { BL [] 0 setdash 2 copy moveto
       2 copy  vpt 180 270 arc closepath fill
               vpt 0 360 arc closepath } bind def
/C5 { BL [] 0 setdash 2 copy moveto
       2 copy  vpt 0 90 arc
       2 copy moveto
       2 copy  vpt 180 270 arc closepath fill
               vpt 0 360 arc } bind def
/C6 { BL [] 0 setdash 2 copy moveto
      2 copy  vpt 90 270 arc closepath fill
              vpt 0 360 arc closepath } bind def
/C7 { BL [] 0 setdash 2 copy moveto
      2 copy  vpt 0 270 arc closepath fill
              vpt 0 360 arc closepath } bind def
/C8 { BL [] 0 setdash 2 copy moveto
      2 copy vpt 270 360 arc closepath fill
              vpt 0 360 arc closepath } bind def
/C9 { BL [] 0 setdash 2 copy moveto
      2 copy  vpt 270 450 arc closepath fill
              vpt 0 360 arc closepath } bind def
/C10 { BL [] 0 setdash 2 copy 2 copy moveto vpt 270 360 arc closepath fill
       2 copy moveto
       2 copy vpt 90 180 arc closepath fill
               vpt 0 360 arc closepath } bind def
/C11 { BL [] 0 setdash 2 copy moveto
       2 copy  vpt 0 180 arc closepath fill
       2 copy moveto
       2 copy  vpt 270 360 arc closepath fill
               vpt 0 360 arc closepath } bind def
/C12 { BL [] 0 setdash 2 copy moveto
       2 copy  vpt 180 360 arc closepath fill
               vpt 0 360 arc closepath } bind def
/C13 { BL [] 0 setdash  2 copy moveto
       2 copy  vpt 0 90 arc closepath fill
       2 copy moveto
       2 copy  vpt 180 360 arc closepath fill
               vpt 0 360 arc closepath } bind def
/C14 { BL [] 0 setdash 2 copy moveto
       2 copy  vpt 90 360 arc closepath fill
               vpt 0 360 arc } bind def
/C15 { BL [] 0 setdash 2 copy vpt 0 360 arc closepath fill
               vpt 0 360 arc closepath } bind def
/Rec   { newpath 4 2 roll moveto 1 index 0 rlineto 0 exch rlineto
       neg 0 rlineto closepath } bind def
/Square { dup Rec } bind def
/Bsquare { vpt sub exch vpt sub exch vpt2 Square } bind def
/S0 { BL [] 0 setdash 2 copy moveto 0 vpt rlineto BL Bsquare } bind def
/S1 { BL [] 0 setdash 2 copy vpt Square fill Bsquare } bind def
/S2 { BL [] 0 setdash 2 copy exch vpt sub exch vpt Square fill Bsquare } bind def
/S3 { BL [] 0 setdash 2 copy exch vpt sub exch vpt2 vpt Rec fill Bsquare } bind def
/S4 { BL [] 0 setdash 2 copy exch vpt sub exch vpt sub vpt Square fill Bsquare } bind def
/S5 { BL [] 0 setdash 2 copy 2 copy vpt Square fill
       exch vpt sub exch vpt sub vpt Square fill Bsquare } bind def
/S6 { BL [] 0 setdash 2 copy exch vpt sub exch vpt sub vpt vpt2 Rec fill Bsquare } bind def
/S7 { BL [] 0 setdash 2 copy exch vpt sub exch vpt sub vpt vpt2 Rec fill
       2 copy vpt Square fill
       Bsquare } bind def
/S8 { BL [] 0 setdash 2 copy vpt sub vpt Square fill Bsquare } bind def
/S9 { BL [] 0 setdash 2 copy vpt sub vpt vpt2 Rec fill Bsquare } bind def
/S10 { BL [] 0 setdash 2 copy vpt sub vpt Square fill 2 copy exch vpt sub exch vpt Square fill
       Bsquare } bind def
/S11 { BL [] 0 setdash 2 copy vpt sub vpt Square fill 2 copy exch vpt sub exch vpt2 vpt Rec fill
       Bsquare } bind def
/S12 { BL [] 0 setdash 2 copy exch vpt sub exch vpt sub vpt2 vpt Rec fill Bsquare } bind def
/S13 { BL [] 0 setdash 2 copy exch vpt sub exch vpt sub vpt2 vpt Rec fill
       2 copy vpt Square fill Bsquare } bind def
/S14 { BL [] 0 setdash 2 copy exch vpt sub exch vpt sub vpt2 vpt Rec fill
       2 copy exch vpt sub exch vpt Square fill Bsquare } bind def
/S15 { BL [] 0 setdash 2 copy Bsquare fill Bsquare } bind def
/D0 { gsave translate 45 rotate 0 0 S0 stroke grestore } bind def
/D1 { gsave translate 45 rotate 0 0 S1 stroke grestore } bind def
/D2 { gsave translate 45 rotate 0 0 S2 stroke grestore } bind def
/D3 { gsave translate 45 rotate 0 0 S3 stroke grestore } bind def
/D4 { gsave translate 45 rotate 0 0 S4 stroke grestore } bind def
/D5 { gsave translate 45 rotate 0 0 S5 stroke grestore } bind def
/D6 { gsave translate 45 rotate 0 0 S6 stroke grestore } bind def
/D7 { gsave translate 45 rotate 0 0 S7 stroke grestore } bind def
/D8 { gsave translate 45 rotate 0 0 S8 stroke grestore } bind def
/D9 { gsave translate 45 rotate 0 0 S9 stroke grestore } bind def
/D10 { gsave translate 45 rotate 0 0 S10 stroke grestore } bind def
/D11 { gsave translate 45 rotate 0 0 S11 stroke grestore } bind def
/D12 { gsave translate 45 rotate 0 0 S12 stroke grestore } bind def
/D13 { gsave translate 45 rotate 0 0 S13 stroke grestore } bind def
/D14 { gsave translate 45 rotate 0 0 S14 stroke grestore } bind def
/D15 { gsave translate 45 rotate 0 0 S15 stroke grestore } bind def
/DiaE { stroke [] 0 setdash vpt add M
  hpt neg vpt neg V hpt vpt neg V
  hpt vpt V hpt neg vpt V closepath stroke } def
/BoxE { stroke [] 0 setdash exch hpt sub exch vpt add M
  0 vpt2 neg V hpt2 0 V 0 vpt2 V
  hpt2 neg 0 V closepath stroke } def
/TriUE { stroke [] 0 setdash vpt 1.12 mul add M
  hpt neg vpt -1.62 mul V
  hpt 2 mul 0 V
  hpt neg vpt 1.62 mul V closepath stroke } def
/TriDE { stroke [] 0 setdash vpt 1.12 mul sub M
  hpt neg vpt 1.62 mul V
  hpt 2 mul 0 V
  hpt neg vpt -1.62 mul V closepath stroke } def
/PentE { stroke [] 0 setdash gsave
  translate 0 hpt M 4 {72 rotate 0 hpt L} repeat
  closepath stroke grestore } def
/CircE { stroke [] 0 setdash 
  hpt 0 360 arc stroke } def
/Opaque { gsave closepath 1 setgray fill grestore 0 setgray closepath } def
/DiaW { stroke [] 0 setdash vpt add M
  hpt neg vpt neg V hpt vpt neg V
  hpt vpt V hpt neg vpt V Opaque stroke } def
/BoxW { stroke [] 0 setdash exch hpt sub exch vpt add M
  0 vpt2 neg V hpt2 0 V 0 vpt2 V
  hpt2 neg 0 V Opaque stroke } def
/TriUW { stroke [] 0 setdash vpt 1.12 mul add M
  hpt neg vpt -1.62 mul V
  hpt 2 mul 0 V
  hpt neg vpt 1.62 mul V Opaque stroke } def
/TriDW { stroke [] 0 setdash vpt 1.12 mul sub M
  hpt neg vpt 1.62 mul V
  hpt 2 mul 0 V
  hpt neg vpt -1.62 mul V Opaque stroke } def
/PentW { stroke [] 0 setdash gsave
  translate 0 hpt M 4 {72 rotate 0 hpt L} repeat
  Opaque stroke grestore } def
/CircW { stroke [] 0 setdash 
  hpt 0 360 arc Opaque stroke } def
/BoxFill { gsave Rec 1 setgray fill grestore } def
end
}}%
\begin{picture}(3600,2160)(0,-50)%
{\GNUPLOTspecial{"
gnudict begin
gsave
0 0 translate
0.100 0.100 scale
0 setgray
newpath
1.000 UL
LTb
525 300 M
0 18 V
0 1742 R
0 -18 V
1175 300 M
0 18 V
0 1742 R
0 -18 V
1825 300 M
0 18 V
0 1742 R
0 -18 V
2475 300 M
0 18 V
0 1742 R
0 -18 V
3125 300 M
0 18 V
0 1742 R
0 -18 V
1.000 UL
LTb
200 300 M
3250 0 V
0 1760 V
-3250 0 V
200 300 L
1.000 UL
LT0
200 301 M
4 0 V
8 -1 V
9 0 V
9 0 V
8 0 V
9 0 V
8 0 V
9 0 V
9 0 V
8 0 V
9 0 V
9 0 V
8 0 V
9 0 V
8 0 V
9 0 V
9 0 V
8 0 V
9 0 V
9 0 V
8 0 V
9 0 V
8 0 V
9 0 V
9 0 V
8 0 V
9 0 V
9 0 V
8 0 V
9 0 V
8 0 V
9 0 V
9 0 V
8 0 V
9 0 V
9 1 V
8 5 V
9 8 V
8 -3 V
9 -8 V
9 -3 V
8 0 V
9 0 V
9 0 V
8 0 V
9 0 V
8 0 V
9 0 V
9 0 V
8 0 V
9 0 V
9 0 V
8 0 V
9 0 V
8 0 V
9 0 V
9 0 V
8 0 V
9 0 V
9 0 V
8 0 V
9 0 V
8 0 V
9 0 V
9 0 V
8 0 V
9 0 V
9 0 V
8 0 V
9 1 V
8 0 V
9 1 V
9 1 V
8 18 V
9 84 V
8 58 V
9 -68 V
9 -78 V
8 -14 V
9 -1 V
9 -1 V
8 -1 V
9 0 V
8 0 V
9 0 V
9 0 V
8 0 V
9 0 V
9 0 V
8 0 V
9 0 V
8 0 V
9 0 V
9 0 V
8 0 V
9 0 V
9 0 V
8 0 V
9 0 V
8 0 V
9 0 V
9 0 V
8 0 V
9 0 V
9 0 V
8 1 V
9 0 V
8 2 V
9 1 V
9 11 V
8 8 V
9 216 V
9 531 V
8 106 V
9 -507 V
8 -325 V
9 -28 V
9 -9 V
8 -4 V
9 -1 V
9 -1 V
8 0 V
9 -1 V
8 0 V
9 0 V
9 0 V
8 0 V
9 0 V
9 0 V
8 0 V
9 0 V
8 0 V
9 0 V
9 0 V
8 0 V
9 0 V
9 0 V
8 0 V
9 0 V
8 0 V
9 0 V
9 1 V
8 0 V
9 0 V
9 2 V
8 2 V
9 7 V
8 12 V
9 72 V
9 643 V
8 898 V
9 -260 V
9 -964 V
8 -376 V
9 -14 V
8 -17 V
9 -1 V
9 -3 V
8 0 V
9 -1 V
9 0 V
8 -1 V
9 0 V
8 0 V
9 0 V
9 0 V
8 0 V
9 0 V
9 0 V
8 0 V
9 0 V
8 0 V
9 0 V
9 0 V
8 0 V
9 0 V
9 0 V
8 0 V
9 0 V
8 0 V
9 0 V
9 0 V
8 0 V
9 0 V
9 0 V
8 1 V
9 6 V
8 18 V
9 11 V
9 -12 V
8 -17 V
9 -6 V
8 -1 V
9 0 V
9 0 V
8 0 V
9 0 V
9 0 V
8 0 V
9 0 V
8 0 V
9 0 V
9 0 V
8 0 V
9 0 V
9 0 V
8 0 V
9 0 V
8 0 V
9 0 V
9 0 V
8 0 V
9 0 V
9 0 V
8 0 V
9 0 V
8 1 V
9 0 V
9 1 V
8 0 V
9 3 V
9 1 V
8 17 V
9 14 V
8 379 V
9 969 V
9 261 V
8 -902 V
9 -647 V
9 -73 V
8 -12 V
9 -7 V
8 -2 V
9 -2 V
9 0 V
8 0 V
9 0 V
9 -1 V
8 0 V
9 0 V
8 0 V
9 0 V
9 0 V
8 0 V
9 0 V
9 0 V
8 0 V
9 0 V
8 0 V
9 0 V
9 0 V
8 0 V
9 0 V
9 0 V
8 0 V
9 1 V
8 0 V
9 1 V
9 1 V
8 4 V
9 9 V
9 30 V
8 338 V
9 533 V
8 -106 V
9 -558 V
9 -229 V
8 -8 V
9 -12 V
9 -1 V
8 -2 V
9 0 V
8 0 V
9 -1 V
9 0 V
8 0 V
9 0 V
9 0 V
8 0 V
9 0 V
8 0 V
9 0 V
9 0 V
8 0 V
9 0 V
9 0 V
8 0 V
9 0 V
8 0 V
9 0 V
9 0 V
8 0 V
9 0 V
9 0 V
8 0 V
9 1 V
8 0 V
9 1 V
9 1 V
8 16 V
9 86 V
9 78 V
8 -65 V
9 -94 V
8 -21 V
9 0 V
9 -2 V
8 0 V
9 -1 V
8 0 V
9 0 V
9 0 V
8 0 V
9 0 V
9 0 V
8 0 V
9 0 V
8 0 V
9 0 V
9 0 V
8 0 V
9 0 V
9 0 V
8 0 V
9 0 V
8 0 V
9 0 V
9 0 V
8 0 V
9 0 V
9 0 V
8 0 V
9 0 V
8 0 V
9 0 V
9 0 V
8 0 V
9 4 V
9 9 V
8 4 V
9 -9 V
8 -7 V
9 -1 V
9 0 V
8 0 V
9 0 V
9 0 V
8 0 V
9 0 V
8 0 V
9 0 V
9 0 V
8 0 V
9 0 V
9 0 V
8 0 V
9 0 V
8 0 V
9 0 V
9 0 V
8 0 V
9 0 V
9 0 V
8 0 V
9 0 V
8 0 V
9 0 V
9 0 V
8 0 V
9 0 V
9 0 V
8 0 V
9 0 V
8 0 V
9 0 V
9 0 V
8 1 V
4 0 V
stroke
grestore
end
showpage
}}%
\put(1825,50){\makebox(0,0){$p_z \left[ \hbar/a_\perp \right]$}}%
\put(100,1180){%
\special{ps: gsave currentpoint currentpoint translate
270 rotate neg exch neg exch translate}%
\makebox(0,0)[b]{\shortstack{$Q \ S(Q,\Omega)/m$}}%
\special{ps: currentpoint grestore moveto}%
}%
\put(3125,200){\makebox(0,0){$4$}}%
\put(2475,200){\makebox(0,0){$2$}}%
\put(1825,200){\makebox(0,0){$0$}}%
\put(1175,200){\makebox(0,0){$-2$}}%
\put(525,200){\makebox(0,0){$-4$}}%
\end{picture}%
\endgroup
 